\documentclass[manuscript]{acmart}
\AtBeginDocument{%
  }

\setcopyright{acmlicensed}
\copyrightyear{2018}
\acmYear{2018}
\acmDOI{XXXXXXX.XXXXXXX}
\acmConference[Conference acronym 'XX]{Make sure to enter the correct
  conference title from your rights confirmation email}{June 03--05,
  2018}{Woodstock, NY}
\acmISBN{978-1-4503-XXXX-X/2018/06}

\usepackage{graphicx} 
\usepackage{url}
\usepackage{xcolor}
\usepackage{amsmath}
\usepackage{booktabs}
\usepackage{tabularx}
\usepackage[most]{tcolorbox}
\usepackage{enumitem}

\definecolor{actblue}{HTML}{729ECE}
\definecolor{actorange}{HTML}{FF9E4A}

\definecolor{actblue}{HTML}{729ECE}
\definecolor{actorange}{HTML}{FF9E4A}

\definecolor{actred}{HTML}{E15759}
\definecolor{actteal}{HTML}{76B7B2}
\definecolor{actgreen}{HTML}{59A14F}
\definecolor{actyellow}{HTML}{EDC948}
\definecolor{actpurple}{HTML}{B07AA1}
\definecolor{actpink}{HTML}{FF9DA7}
\definecolor{actbrown}{HTML}{9C755F}

\definecolor{actreport}{HTML}{4E79A7}   
\definecolor{actanswer}{HTML}{ff7f0f}   

\newcommand{\actbox}[2]{%
  \tcbox[
    on line,
    boxsep=0pt,
    left=2.5pt,
    right=2.5pt,
    top=0.8pt,
    bottom=0.8pt,
    colframe=#1!75!white,
    colback=#1!40!white,
    boxrule=0.4pt,
    arc=2pt,
    nobeforeafter
  ]{\ttfamily\small #2}}

\newcommand{\actTA}{\actbox{actblue}{Think Aloud}}
\newcommand{\actSR}{\actbox{actorange}{Solution Request}}
\newcommand{\actCGQ}{\actbox{actteal}{Common Ground Question}}
\newcommand{\actKDQ}{\actbox{actpurple}{Knowledge Deficit Question}}
\newcommand{\actCA}{\actbox{actgreen}{Conversational Acknowledgment}}
\newcommand{\actMC}{\actbox{actbrown}{Metacomment}}

\newcommand{\actBR}{\actbox{actreport}{Board Report}}
\newcommand{\actPCA}{\actbox{actanswer}{Provide Correct Answer}}
\newcommand{\actGP}{\actbox{actyellow}{General Principle}}
\newcommand{\actPrompt}{\actbox{actred}{Prompt}}
\newcommand{\actHint}{\actbox{actpink}{Hint}}

\begin{document}


\title[Human-AI Conversational Behaviors Predict Unassisted Task Performance]{Human-AI Conversational Behaviors Predict Unassisted Task Performance}

\author{Li Siyan}
\correspondingauthor
\authornote{Work done during internship at Together AI.}
\email{siyan.li@columbia.edu}
\orcid{0009-0002-1465-6337}

\affiliation{%
  \institution{Columbia University}
  \city{New York}
  \state{New York}
  \country{USA}
}

\author{Federico Bianchi}
\affiliation{%
  \institution{Together AI}
  \city{San Francisco}
  \state{California}
  \country{USA}
}

\author{James Zou}
\affiliation{%
  \institution{Stanford University}
  \city{Stanford}
  \state{California}
  \country{USA}
}
\affiliation{%
  \institution{Together AI}
  \city{San Francisco}
  \state{California}
  \country{USA}
}

\author{Kaitlyn Zhou}
\affiliation{%
  \institution{Cornell University}
  \city{Ithaca}
  \state{New York}
  \country{USA}
}
\affiliation{%
  \institution{Together AI}
  \city{San Francisco}
  \state{California}
  \country{USA}
}

\renewcommand{\shortauthors}{Trovato et al.}

\begin{abstract}
    Assistance from AI tools has supported and improved human performance across domains. However, recent research suggests that these immediate benefits may entail future costs, including diminished performance when AI assistance is no longer available. We study how human-AI interaction behaviors correlate with immediate and future unassisted task performance across two game-based problem-solving user studies ($n=139$ and $n=111$), using a \textit{dialogue act} framework adopted from a tutor-student dialogue taxonomy. In our studies, verbalizing thought processes correlates with higher unassisted outcomes, whereas directly requesting solutions correlates negatively. Similar to these participant-side patterns, assistant explanations of the current problem state are associated with better subsequent unassisted performance, whereas directly providing the next action is associated with worse performance. Qualitative and subtype analyses further show that ostensibly similar reasoning turns can elicit different assistance. Our findings suggest that preserving users' cognitive participation in problem-solving may support performance beyond AI-assisted interaction.
\end{abstract}



\received{20 February 2007}
\received[revised]{12 March 2009}
\received[accepted]{5 June 2009}

\maketitle

\begin{figure}
  \includegraphics[width=\textwidth]{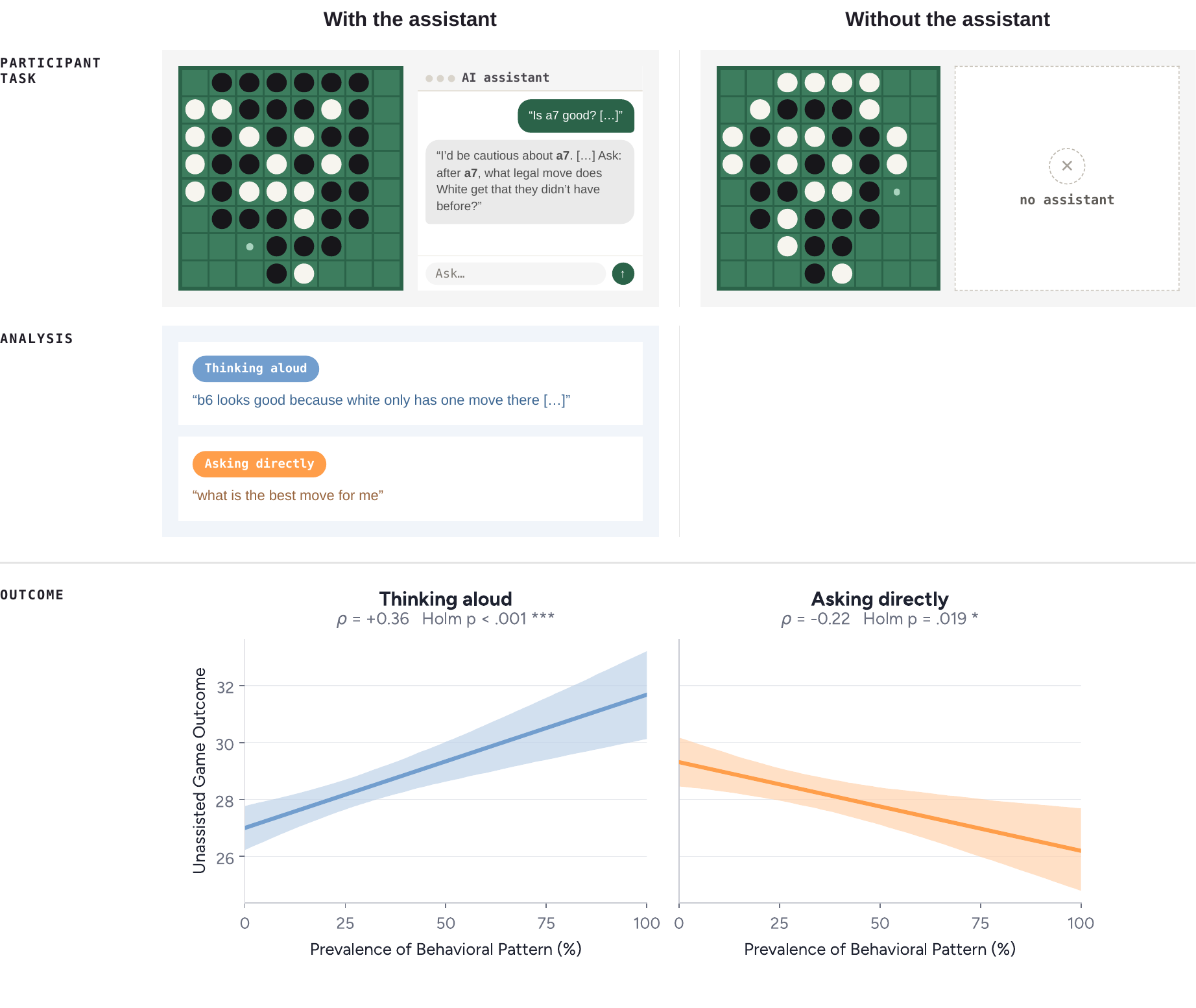}
  \caption{How participants talked to the AI assistant correlates with how well they played
  without it. Each participant solved one game puzzle with a chat assistant available and two related puzzles without it. We analyze participant conversations with the assistant and identify two behaviors: thinking aloud, which is positively correlated with unassisted performance, and asking directly, which is negatively correlated with it. These are observational associations, not effects of the behaviors.}
  \Description{A three-row, two-column diagram. The columns are labelled ``With the assistant'' and ``Without the assistant''. Top row, Participant task: on the left, an Othello board in mid-endgame beside a chat window in which a participant asks the assistant about a candidate square; on the right, the same board is greyed out beside an empty panel marked ``no assistant''. Middle row, Analysis: two cards appear under the assisted column only, one tagged ``Thinking aloud'' in blue, quoting a participant reasoning aloud about a move, the other tagged ``Asking directly'' in orange quoting another participant asking for the best move. The unassisted column is empty here because there is no conversation to analyze. Bottom row, Outcome: two line charts spanning both columns. In each, the horizontal axis is Prevalence of Behavioral Pattern from 0 to 100 percent, and the shared vertical axis is the participant's outcome. In the left chart, ``Thinking aloud'', the fitted line rises from 27.0 at 0 percent prevalence to 31.7 at 100 percent, Spearman rho plus 0.36, Holm-corrected p below .001. In the right chart, ``Asking directly'', the line falls from 29.3 to 26.2, rho minus 0.22, p equals .019. Each line carries a shaded 95 percent bootstrap band that widens toward the right-hand end. 111 participants.}
  \label{fig:teaser}
\end{figure}

\section{Introduction}

Assistance from AI tools can substantially boost user performance while assistance is available, improving both output quality and efficiency across tasks \cite{noy2023experimental,brynjolfsson2025generative,vaccaro2024combinations,choi2024lawyering,peng2023impact}.
However, recent work shows that these immediate gains may carry long-term costs. Growing evidence links AI over-reliance \cite{passi2022overreliance} to cognitive activity decline \cite{kosmyna2025your}, reduced perceived idea ownership and confidence \cite{baldeo2026generative,lee2025impact}, lower memory accuracy for human-AI collaboration artifacts \cite{zindulka2026ai}, and worse exam scores despite more efficient homework completion \cite{stromberg2026generative}. Yet as AI becomes increasingly embedded in education and knowledge work \cite{dell2023navigating,baidoo2023education,ellingrud2023generative,dong2025examining,ciampa2023chatgpt}, abandoning the technology altogether is neither realistic nor desirable. The challenge, then, is not whether to use AI, but how to enable interactions that preserve human agency, learning, and critical engagement while still capturing its benefits.

This distinction is increasingly reflected in research on AI-supported learning. Unconstrained AI access may improve performance during practice while hindering unassisted outcome \cite{stromberg2026generative}; in contrast, installing pedagogical guardrails reduces damage to independent learning \cite{bastani2025generative}. Other work similarly finds benefits from pedagogically-informed AI systems that promote generative sense-making, solicit explanations, or proactively guide learners rather than only providing solutions \cite{makransky2025wow}. Engaging general AI assistants through Socratic dialogue has likewise been linked to better learning \cite{mcnichols2026studychat}. Together, these results suggest that learning with AI depends not only on the information an assistant provides, but on how reasoning work is distributed between the user and the assistant.
This echoes longstanding literature on self-explanation: students benefit when they actively articulate and reason through problems themselves \cite{graesser2005autotutor,chi1994eliciting,chi2014icap}. Similarly, effective tutoring systems often scaffold students through questions, explanations, and hints instead of simply supplying answers \cite{winkler2020sara,ma2014intelligent,scarlatos2025training,latif2026systematic}. 
In these systems, however, the pedagogical policy is typically imposed in a top-down manner, with education experts involved in the design process. General-purpose AI assistants allow substantially more flexibility, as they impose fewer pedagogical constraints \cite{team2024learnlm}: the same assistant can act as an explainer, evaluator, tutor, or solution provider depending in part on how the user engages with it.

This motivates a complementary question to prior work on system-imposed scaffolding: \textit{when users interact freely with an assistant that permits multiple modes of assistance, which conversational behaviors track the retention of performance when AI support is no longer available?}


We take an interaction-centered, bottom-up approach, examining the conversational actions that emerge when users freely interact with a less pedagogically constrained AI assistant. This perspective draws on pragmatics and speech-act literature, which treats utterances as action units during communication in addition to carriers of information \cite{austin1975things,searle1969speech}. Dialogue acts provide a structured representation of the communicative function of user utterances, allowing researchers to aggregate, compare, and analyze recurring conversational behaviors and interaction patterns \cite{stolcke2000dialogue,d2010mining,anderson1991hcrc,welivita2020taxonomy}. This lens lets us examine \textit{how} conversational moves distribute reasoning work between user and assistant, and which interaction patterns correspond to later independent performance.

In this work, we focus on understanding the human-AI interactions that best support \textit{unassisted near-transfer}, in which unassisted performance is measured on unseen, \textbf{similar} tasks \cite{perkins1992transfer}, as opposed to far-transfer, where learned skills are applied to radically different domains. For example, after solving board positions with AI support, participants in our studies must independently solve a new board position with similar motifs from the same type of game. Specifically, our research aims to answer the following questions:
\begin{enumerate}
    \item Which participant conversational behaviors are associated with subsequent unassisted near-transfer performance?
    \item How do participant behaviors shape assistant responses, and which assistant response behaviors are associated with subsequent unassisted near-transfer performance?
\end{enumerate}

Across two studies, we find that articulating one's reasoning correlates positively with unassisted performance ($\rho = +0.36, p<.001$), whereas directly requesting solutions correlates negatively ($\rho = -0.22, p=.019$). Mirroring these participant-side results, assistant explanations of the current problem state show a positive relationship with subsequent unassisted performance ($\rho = +0.227, p=.017$), whereas providing the next action shows a negative one ($\rho = -0.243, p=.010$). We then present qualitative case studies of conversations with participants who differ in their assisted and transfer performance, revealing the nuanced subcategories of the \textit{thinking aloud} speech act.

\textbf{Our work offers three core contributions.} First, we connect conversational behaviors during AI-supported problem-solving with later near-transfer performance without assistance, linking patterns in human–AI interaction to educational theories of self-explanation and help-seeking. Those theories were developed for students working with instructional material or with tutors that constrain what help is given. We show that the same distinctions hold for adults using a general-purpose assistant that supplies answers on request. 
Second, we show that users' conversational behavior is systematically related to the form of AI assistance they receive, highlighting dialogue as both a trace of cognitive participation and a means by which users shape assistance.
Third, we provide both qualitative and quantitative subtype analyses that offer a fine-grained view of different forms of the \textit{thinking aloud} dialogue act, demonstrating that it is not homogeneous: utterances given the same label can prompt meaningfully different assistant behaviors. We outline design implications and possible future learning scaffolds that could teach less proficient AI users interaction strategies that best support unassisted near-transfer. Overall, our findings deepen understanding of what may enable effective AI use going forward and what may need to be built in to promote these beneficial behaviors in future interactions.


%




\section{Related Work}





\subsection{Conversational Behavior and Reliance in Human-LLM Interaction}

User idiosyncrasies -- request styles, monitoring efforts, and delegation habits -- are reflected through their interactions with LLMs, which in turn influence task success and conversation dynamics. Recent work examines how conversational behaviors shape human-AI collaboration outcomes, measured by user satisfaction \cite{kim2024understanding} and objective task success \cite{xiao2024devgpt,awasthi2026turnstyle}, while broader analyses characterize recurring roles and divisions of responsibility between humans and LLMs \cite{chappidi2026does,zou2026llm}. How users position themselves relative to the model, these findings suggest, shapes the collaboration.

A particularly important dimension of these interactions is how much cognitive labor is delegated to AI. Over-reliance and over-delegation, often shaped by AI-related attitudes and beliefs \cite{klingbeil2024trust, biswas2026belief}, can lead users to accept incorrect LLM outputs without critical examination \cite{zhai2024effects,spatharioti2025effects}, and behavioral traces such as copy-pasting, limited task comprehension, and accepting information despite hesitation can reveal such reliance during conversation \cite{liu2026behavioral}. Crucially, cognitive forcing functions can reduce it \cite{buccinca2021trust}. Reliance is thus enacted through the interaction itself, as users continuously decide what reasoning to retain and what work to delegate.

A smaller body of work connects these dynamics to outcomes beyond immediate metrics. \citeauthor{furniturewala2026learning} analyze human-LLM conversations on socio-political issues, finding that changes in political knowledge and self-efficacy do not emerge from LLM explanation quality alone but rest upon interactions between engagement quality and user traits \cite{furniturewala2026learning}; in StudyChat \cite{mcnichols2026studychat} and \citeauthor{brender2026reflective} \cite{brender2026reflective}, conceptually oriented engagement was linked to better outcomes than solution-oriented use. Nevertheless, most conversational analyses focus on immediately measurable outcomes of assisted tasks, including task success, reliance, and perceived helpfulness. Work that does measure subsequent transfer has also largely examined students in educational settings. This leaves open the question of how naturally emerging interaction strategies relate to retained performance among broader user populations. We address this gap by examining how conversational behaviors during AI-assisted problem solving relate to what users can subsequently accomplish \textit{without} AI assistance, and how those same behaviors shape the forms of assistance they receive.

\subsection{Cognitive Engagement and Help-Seeking in AI-Assisted Learning}


Learning outcomes have been empirically linked to learners' cognitive engagement with the material. The ICAP framework \cite{chi2014icap} specifies four modes of engagement -- Interactive, Constructive, Active, and Passive -- with deeper learning attainable as learners progress from Passive to Interactive. One well-established form of constructive learning is self-explanation: eliciting learners to explain their own reasoning can improve understanding and assist transfer \cite{chi1994eliciting,vanlehn1992model,rittle2017eliciting}, and a meta-analysis of 64 studies finds that prompting self-explanation yields moderate learning improvements across instructional contexts, with learner-generated explanations sometimes outperforming instructional ones \cite{bisra2018inducing}.

A complementary literature examines how learners seek assistance, distinguishing \textit{instrumental help-seeking}, in which learners request only enough information to continue solving a problem themselves, from \textit{executive help-seeking}, in which they seek to have the problem or next step completed for them \cite{nelson1981help,karabenick2004perceived,aleven2003help}. Although executive help can facilitate immediate task completion, it accompanies poorer subsequent learning \cite{aleven2003help,karabenick2003seeking}. Intelligent tutoring systems are therefore often designed to preserve cognitive engagement \cite{kestin2025ai,pardos2024chatgpt}: Cognitive Tutors that prompt students to explain their problem-solving steps yield greater conceptual understanding and transfer than otherwise similar tutoring \cite{aleven2002effective}, self-explanation of worked examples in algebra tutors can further improve conceptual and procedural learning \cite{booth2013using}, and conversational tutors use prompts, hints, and scaffolding to guide learners toward constructing answers themselves \cite{graesser2005autotutor,winkler2020sara}.

The same distinction appears to matter when learners interact with LLMs. Complementing StudyChat \cite{mcnichols2026studychat}, \citeauthor{bastani2025generative} show that unrestricted access to a general-purpose LLM can improve performance while assistance is available yet impair subsequent unassisted performance, whereas pedagogical assistance that preserves opportunities for cognitive engagement avoids much of this harm \cite{bastani2025generative}. The educational consequences of AI assistance therefore depend on whether the interaction supports or substitutes for learners' own cognitive work.






\section{Methods: User Study Design}
\label{sec:methods}

\begin{table}[t]
    \centering
    \small
    \begin{tabularx}{\linewidth}{@{}lXX@{}}
    \toprule
      & \textbf{Connect Four} & \textbf{Othello} \\
    \midrule
      Role & Discovery & Replication \\
      Puzzles & $CF_{A}$ assisted, then $CF_{U}$ & $O_{A}$ assisted, then $O_{U1}$, $O_{U2}$ \\
      Conversation & No minimum & At least five participant turns \\
      Conditions ($n$) & Standard-AI (139) & Standard-AI (111); Baseline, no AI (100) \\
      Outcome & Board Score on $CF_{U}$: optimal moves, 0--3 & Disc count of 64, averaged over $O_{U1}$, $O_{U2}$ \\
    \bottomrule
    \end{tabularx}
    \caption{Our two game-based user studies. See more details in Section \ref{sec:methods}.}
    \label{tab:study-comparison}
\end{table}

Across our between-subject studies (Table \ref{tab:study-comparison}), participants completed game-based problem-solving tasks both with access to an AI assistant and independently. The two studies share common components and design considerations: task design desiderata, prompting-proficiency screener, AI assistant, and procedure. As we focus on near-transfer as the outcome, the assisted task should precede the unassisted task, and they must be similar, which warrants deliberate curation and design (Section \ref{sec:task_design_des}). We included the prompting-proficiency screener (Section \ref{sec:prompt_creation_screener}) as a quality control for participants: they should ideally possess the required skills to interact with the assistant productively. During the assisted task, the AI assistance was engineered to improve experimental control and avoid confounding our results with variations in assistant response correctness (Section \ref{sec:ai_assistant_design}). The full study procedure included additional surveys to assess engagement and experiences (see Section \ref{sec:study_procedure}). Experimental differences are described in their respective sections.

\subsection{Task Design Desiderata}
\label{sec:task_design_des}

Our goal is to study how conversational behavior during AI-assisted problem solving relates to subsequent performance without AI. This required a task environment that was both experimentally controlled and rich and challenging enough to elicit meaningful interaction. Prior work has used game-like settings to promote engagement and high-quality participation in online tasks \cite{krause2015play,eickhoff2012quality}, making games a useful setting for studying repeated task performance. In particular, we sought game tasks that participants could learn quickly, limiting the influence of specialized domain knowledge; that were sufficiently challenging to create a genuine need for AI assistance; and whose outcomes could be scored objectively against a known solution. The tasks also needed to support closely matched, unseen variants so that we could measure near-transfer after assistance was removed. Finally, because our analysis depends on participants meaningfully engaging with the assistant, we favored interactive, interesting tasks over purely mechanical ones.


We operationalized these requirements through games, specifically Connect Four and Othello. They are popular games with simple rules, yet still challenging because of the sheer number of possible board positions \cite{uiterwijk1989knowledge,iwata1994othello}. Since they are both solved games (i.e., the optimal move can be calculated), the quality of each move can be objectively assessed. At the same time, different board positions can represent similar strategic motifs, allowing us to select assisted and unassisted puzzles that require related reasoning without sharing the same solution. We used end-game puzzles rather than complete games to standardize task duration and reduce variation across participants. Participants played a fixed number of moves in Connect Four puzzles, but we let them play Othello until the end because final performance in Othello is more difficult to estimate for beginners.


To assess \textit{unassisted near-transfer}, we selected subsequent puzzles that shared \textit{strategic motifs} with the initial puzzle but had different initial optimal moves. E.g., for Connect Four, both puzzles ($CF_{A}$, $CF_{U}$) require reasoning about play in the central columns and for Othello ($O_{A}$, $O_{U1}$, $O_{U2}$), all puzzles involve strategic play around X-squares and avoiding tempting but strategically suboptimal greedy moves.

We pre-computed optimal strategies for the opponent using Minimax \cite{v1928theorie}. During each game, the computer opponent followed the corresponding optimal strategy. Pre-computation both reduced system latency and provided a consistent basis for evaluating participant performance.

Using gameplay simulations, we scored each participant based on the number of optimal moves they made given the state of the board (\textbf{Board Score}). For Othello, we focus onthe \textbf{disc count} metric, the number of discs of a specific color on the board; a participant with more than 32 discs wins for our specific Othello board size ($8 \times 8$).

\subsection{Prompt Creation Screener}
\label{sec:prompt_creation_screener}

We targeted participants who were sufficiently competent at interacting with AI in order to maximize participants' chances of utilizing their AI interactions to boost subsequent unassisted performance. 
Because inquiring about usage frequency prior to the study can bias participants, and Prolific's pre-screening can be incomplete, we instead used \textbf{prompt-construction ability} as a proxy for competency level. Prior work has shown that AI usage does not directly predict AI literacy \cite{bewersdorff2025ai}, and that self-assessment of AI prompting skills may be inaccurate \cite{zhang2026assess}. Participants completed a prompt-creation task based on a fictional scenario (Appendix~\ref{app:prompt_creation}) in which they were asked to formulate an instruction for an AI assistant. Those who scored at or above the screener threshold (median score 3 out of 4) qualified.  

Prompts were automatically evaluated by a panel of LLM-based judges. We validated these scores against human annotations and observed substantial agreement between the panel's mean score and the human's (ICC(2,1) $= 0.787$; Pearson's $r = 0.829$, $p < .001$; Appendix~\ref{app:prompt_creation}).

\subsection{AI Assistant Design}
\label{sec:ai_assistant_design}

We designed the AI assistant to isolate variation in the form of assistance from variation in its underlying solution quality. We instantiated our AI assistant as \texttt{GPT-5.5} with low reasoning effort to leverage a high-capability model while minimizing latency. We prompted the assistant to act as a coach and included the primary gameplay principles in the prompt, while explicitly permitting it to supply answers directly on request (Appendix~\ref{app:prompts}). The assistant was therefore less pedagogically constrained than a tutoring system with a top-down pedagogical policy. 

To reduce potential errors in the assistant's responses, we provided the assistant with \textbf{the participant's next optimal move} and an \textbf{image of the board}. We made this design choice deliberately to reduce variation in assistant correctness, but the assistant could still deliver the same fixed solutions in different ways depending on participants' conversation strategies. Holding the underlying solution constant, therefore, allows us to study how conversational behavior shapes the assistance users receive and how those interaction patterns relate to subsequent unassisted performance.

\subsection{Study Procedure}
\label{sec:study_procedure}

All game-based studies followed a common procedure.

\begin{enumerate}
\item \textbf{Consent and initial eligibility:}
Participants first provided informed consent (Appendix~\ref{app:informed_consent}) and reported their prior experience with the assigned game. Participants who selected ``Agree'' or ``Strongly Agree'' in response to ``I am a skilled [Game] player'' were excluded.

\item \textbf{Prompting proficiency screener:}
Participants completed the prompt-creation task described above, which we used to characterize their prompt-construction proficiency.

\item \textbf{Initial puzzle:}
\textit{Participants started the round without AI assistance}. After their first move, participants rated their move confidence. As an additional check against substantial prior game expertise, participants who both selected the correct first move and reported confidence above ``Moderately confident'' were excluded. Because this first move is made before the assistant becomes available, its correctness and confidence rating, together with the self-rated skill item from step 1, are the only measures of game ability we hold that precede any AI interaction; we use them as pre-assistance covariates (Sections~\ref{sec:cf-prior-skill} and~\ref{sec:othello-focal}). Participants completed the first puzzle task under the assistance condition assigned in the corresponding study. Depending on the condition, participants either completed the puzzle independently or could converse with the AI assistant.

\item \textbf{Post-task questionnaire:}
Participants completed the NASA-TLX and, where applicable, reported their perceptions of the AI assistant and reflected on their interaction with it. Participants did not receive feedback about whether their moves were optimal until after completing the corresponding NASA-TLX questionnaire. We delayed this feedback to avoid influencing participants' retrospective assessments of workload and performance. We administered the AI-assistant questionnaire through a separate Google Form opened in a new browser tab, physically separating evaluation of the assistant from the conversational interface to reduce potential media-equation effects \cite{reeves1996media}.

\item \textbf{Unassisted near-transfer puzzles:}
Participants then completed the remaining end-game puzzles without access to the AI assistant. Connect Four participants completed one additional puzzle, whereas Othello participants completed two. Performance on these puzzles constitutes our measure of subsequent unassisted near-transfer.

\item \textbf{Final questionnaire:}
After completing the unassisted puzzles, participants completed a final NASA-TLX questionnaire.

\item \textbf{Demographics:}
Finally, participants reported their age, highest level of education, occupation, and frequency of generative AI use. AI usage frequency was measured on an eight-point scale ranging from 0 (``Never'') to 7 (``Several times a day'').

\end{enumerate}

\section{Discovery Study Results: Connect Four}

\subsection{Recruitment}

We recruited participants from Prolific’s “Qualified AI Taskers” pool and required a score $\geq 3$ on the prompt-creation screener. We recruited 270 participants, of whom 165 completed the study. All participants provided informed consent to participate in the study. They were compensated 15 USD per hr, and the study protocol was approved by an external IRB. Because our conversational measures are undefined for participants who never interacted with the assistant, we excluded 26 participants who produced no conversational turns, leaving an analytic sample of \(n=139\).

\subsection{Participant Partition}
We divided participants into two groups using their Board Scores on $CF_{U}$ (number of optimal moves, ranging from 0 to 3 here):
\begin{enumerate}
    \item \textbf{Solvers} (n=79): Participants who made at least two (majority) optimal moves.
    \item \textbf{Strugglers} (n=60): All other participants. 
\end{enumerate}

We chose this binary partition because the outcome variable has limited granularity (ranging from 0 to 3). Instead, we contrasted differences between these two groups to establish preliminary trends to focus on in later studies.

\subsection{Participant Dialogue Act Taxonomy}
\label{sec:taxonomy}

The process of participants consulting the AI assistant can be viewed as a tutoring session. Inspired by education literature, we formulated a preliminary taxonomy for participant dialogue acts using a modified version of the student dialogue acts from \cite{d2010mining}; we added the ``\actSR{}'' dialogue act, where participants ask the AI assistant for solution suggestions (e.g., \textit{``What is my best move here?''}). Each utterance could be assigned multiple acts. Annotations were produced by per-act majority vote of a three-model LLM panel; the panel shares two of its three models with the prompt-quality judges and is specified in Appendix~\ref{app:prompts}. Against labels from one of the researchers on 60 Connect Four utterances, the panel reached Cohen's $\kappa = 0.767$ (substantial agreement). By act, $\kappa = 0.75$ for \actTA{} and $0.93$ for \actSR{}; \actKDQ{} was lowest at $0.56$. The three panel models also agree with one another on every turn in both studies. Fleiss' $\kappa$ is $0.81$ (Connect Four) and $0.75$ (Othello) for \actTA{}, and $0.94$ and $0.86$ for \actSR{} (Appendix~\ref{app:reliability}). On a second, blind sample of 150 turns drawn from both studies (Appendix~\ref{sec:taxonomy-human}), agreement with a human coder was $\kappa = 0.62$ for \actTA{} and $0.94$ for \actSR{}. The annotator was given fourteen candidate acts, of which the six in Table~\ref{tab:taxonomy} occurred in our transcripts. We include the full taxonomy of participant utterances, with how often each act appears, in Table \ref{tab:taxonomy}.

\begin{table}[t]
    \centering
    \small
    \begin{tabularx}{\linewidth}{@{}l c c X@{}}
    \toprule
       \textbf{Dialogue Act} & \textbf{\% of turns} & \textbf{\% of turns} & \textbf{Examples}  \\
        &  Connect Four  & Othello & \\
    \midrule
       \actTA{} & 36 & 31
         & 
           \textit{``I'm thinking 5, it sets up nicely for future opportunities, diagonally''} \newline
           \textit{``If I go there, they can block my existing diagonal three, so I will avoid that.''} \\
    \addlinespace
       \actCGQ{} & 38 & 38
         & \textit{``Is column 3 good?''} \newline
           \textit{``Should I go 2?''}  \\
    \addlinespace
       \actSR{} & 38 & 28
         & \textit{``What should I do now?''} \newline
           \textit{``Where should I put it''} \\
    \addlinespace
       \actCA{} & 13 & 17
         & \textit{``Thank you''} \newline
           \textit{``I will go with it!''} \\
    \addlinespace
       \actKDQ{} & 8 & 17
         &
           \textit{``are there any threats on the board?''} \newline
           \textit{``Can I only go next to another playing piece?''} \\
    \addlinespace
       \actMC{} & 5 & 10
         & \textit{``I don't quite follow (I've never played this game before)''} \newline
           \textit{``I lost comprehensively, following your advice''} \\
    \bottomrule
    \end{tabularx}
    \caption{Our taxonomy for annotating participant utterances in our tasks, modified from \cite{d2010mining}. Percentages are the share of participant turns carrying each act, in Connect Four ($n=139$, 384 turns) and Othello ($n=111$, 616 turns); a turn may carry several acts, so columns do not sum to 100.}
    \label{tab:taxonomy}
\end{table}

\subsection{Thinking Aloud Is Associated with Better Unassisted Transfer}

We examined differences in dynamics with the AI assistant between participants who attained good unassisted performance and those who did not. We measured \textbf{dialogue act prevalence} of each dialogue act, defined as the percentage of a participant's turns that contain an act:
\begin{align*}
    DAP(A, P) &= \frac{\#\text{Turns containing Dialogue Act}\ A}{\#\text{Turns by Participant P}} \times 100
\end{align*}

\begin{figure}
    \centering
    \includegraphics[width=\linewidth]{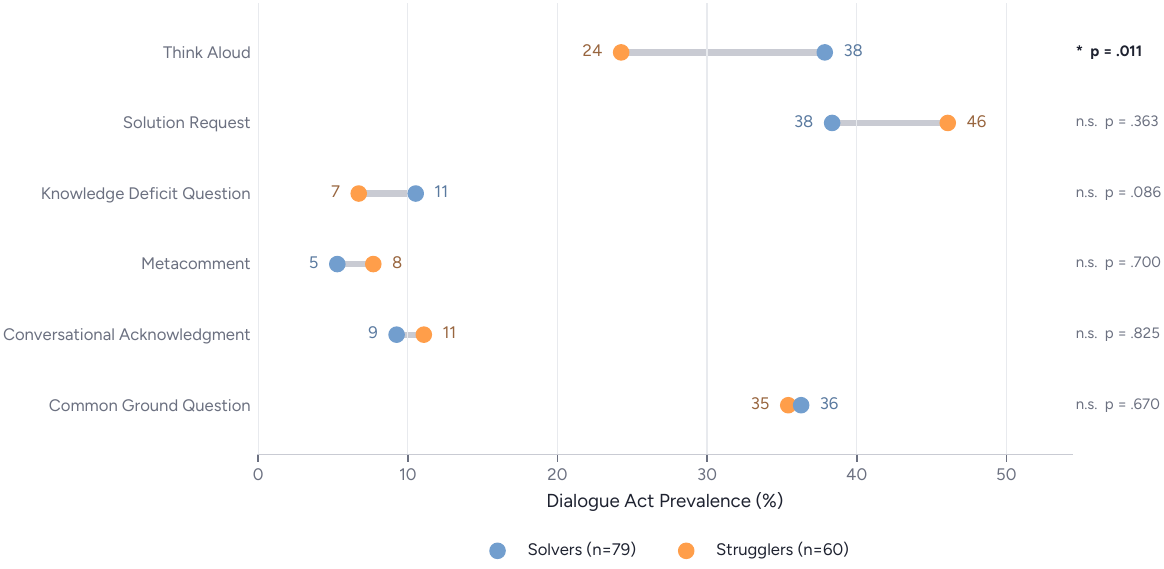}
    \caption{Dumbbell chart of averaged dialogue act prevalence per outcome-based participant group in the Connect Four task.}
    \Description{A dumbbell chart with one row per dialogue act. Each row carries two dots joined by a bar: a blue dot for Solvers ($n = 79$) and an orange dot for Strugglers ($n = 60$), positioned by the mean share of that participant's turns carrying the act, on a horizontal axis running from 0 to about 50 percent. Reading down the rows: Think Aloud, Solvers 38 percent against Strugglers 24 percent, the one significant difference at $p = .011$; Solution Request, 38 against 46, not significant at $p = .363$; Knowledge Deficit Question, 11 against 7, $p = .086$; Metacomment, 5 against 8, $p = .700$; Conversational Acknowledgment, 9 against 11, $p = .825$; and Common Ground Question, 36 against 35, $p = .670$. Each row's p-value is printed at the right-hand edge.}
    \label{fig:dialogue_act_diff_cf}
\end{figure}

The clearest difference between Solvers and Strugglers was the prevalence of \actTA{}: \textbf{Solvers more frequently articulated their thought process compared to Strugglers} (Mann-Whitney $U = 2946.5$, $p = .011$; Figure \ref{fig:dialogue_act_diff_cf}). For instance, one participant categorized as Solver wrote ``im considering placing my next red at 3,3'', while a Struggler participant asked ``what is my next best move''. As this is a discovery study, we compared all dialogue acts between Solvers and Strugglers without a prespecified hypothesis, so the comparisons remain exploratory and are not corrected for multiplicity. 

We highlight that \actSR{} is not incompatible with higher unassisted performance. In fact, Solvers also sought solution help, but they often provided reasoning alongside their requests, for example: ``Where is the best next slot for me to put my red counter in to ensure that I can win. I can see a couple of risks, particularly on the left hand side of the board''. \label{sec:cf-cooccurrence}
To quantify this, we computed the \textbf{reasoning-request co-occurrence rate}: the percentage of solution requests where the request is coupled with thinking aloud. We found that this co-occurrence rate was \textbf{8.6\% for Solvers and only 1.4\% for Strugglers} (Mann-Whitney $U = 2742.0$, $p = .005$). This result suggests that the distinction between the groups may concern not simply whether participants ask for assistance, but whether they contribute their own reasoning when doing so.

We next examined how a participant's prevalence of \actTA{} related to their $CF_{U}$ performance. Because $CF_{U}$'s Board Score is a discrete variable ranging from 0 to 3, we chose proportional-odds ordinal logistic regression to model it. We added $CF_{A}$ scores and AI usage frequency modeled as ordinal covariates. Among the set of dialogue acts, the model containing \actTA{} DAP with all covariates yielded the best BIC values. We report results for three models with progressively more covariates (Table \ref{tab:logistic_regression_cf}).

\begin{table}[]
    \centering
    \begin{tabular}{c|ccc}
    \toprule
       \textbf{Model} & $\beta_{\text{TA-DAP}}$ & $\beta_{\text{AI}}$ & $\beta_{CF_{A}}$ \\
       \midrule
        + TA-DAP & +0.306* & -- &  -- \\
        + AI & +0.336* & -0.198* & -- \\
        + $CF_{A}$ Score & +0.344* & -0.207* & +0.284 \\
        + Pre-assistance skill$^{\dagger}$ & +0.380* & -0.205* & +0.148 \\
        \bottomrule
    \end{tabular}
    \caption{Ordinal logistic regression results, modeling $CF_{U}$ scores as the dependent variable, with different combinations of covariates. \actTA{} DAP is standardized, so $\beta_{\text{TA-DAP}}$ is the log-odds shift per 1 SD; AI usage and $CF_{A}$ score are on their original scales. $^{\dagger}$Adds, to the previous model, three measures taken before the assistant was available: whether the unaided first move was optimal ($\beta = +0.562$, $p = .093$), confidence in that move ($-0.233$, $p = .22$), and self-rated game skill ($+0.161$, $p = .50$). *$p < .05$}
    \label{tab:logistic_regression_cf}
\end{table}

\actTA{} DAP remains positively associated with $CF_{U}$ performance, even when controlling for $CF_{A}$ performance and AI usage frequency ($\beta = +0.344$, $p = .028$). This strengthens the evidence for an association between \actTA{} prevalence and unassisted performance.

\label{sec:cf-prior-skill}
Our results thus far cannot separate prior ability from what the conversation contributed. We therefore added the three pre-assistance measures described in Section~\ref{sec:study_procedure}: correctness of the unaided first move, confidence in it, and self-rated game skill. \actTA{} DAP was unrelated to all three (Spearman $|\rho| \le 0.14$, all $p \ge .10$), and its coefficient was unchanged by them ($\beta = +0.380$, $p = .018$; Table~\ref{tab:logistic_regression_cf}). The adjustment therefore rules out confounding by entry-level skill \emph{as these items measure it}. It does not rule out a more general disposition to engage with the board.


\section{Replication Study Results: Othello}

The Connect Four study provides some initial insights into how users' behavioral patterns during human-AI conversations can relate to their unassisted performance. We examined whether these results would hold in a second game-based task and addressed some limitations in the discovery study. First, most participants engaged with the assistant only briefly, limiting the available signals about interaction patterns. Secondly, unassisted performance was measured using only one puzzle with discrete, narrow score ranges, undermining the robustness of our correlation results. Third, we lacked a completely unassisted baseline to establish how unassisted performance after AI assistance compares with that without it.

Othello directly addresses the low-resolution outcome space of Connect Four by offering a wider range of possible game states and final scores. Played on an $8 \times 8$ grid, two players take turns placing black and white discs. Whenever a newly placed disc traps a continuous line of the opponent’s discs between two discs of the player’s own color, all of the trapped discs are flipped to that player’s color. The game ends when no legal moves remain, and the player with the most discs of their color on the board wins. For outcome, we measured the \textbf{disc count} metric, which is the number of Black pieces on the board (participants all played Black). Every game played to the end filled all 64 squares, so a count above 32 is a win, 32 is a draw, and the count ranges from 0 to 64. For the two unassisted puzzles, we average the two counts.

To address the aforementioned limitations, we therefore required at least five participant turns with the assistant during the assisted Othello round, measured transfer across two subsequent unassisted puzzles, and added a baseline condition in which participants completed all three puzzles without AI assistance.

\subsection{Relationships between Conversational Behaviors and Unassisted Performance Replicate in Othello}
\label{sec:othello-focal}

\begin{figure}
    \centering
    \includegraphics[width=\linewidth]{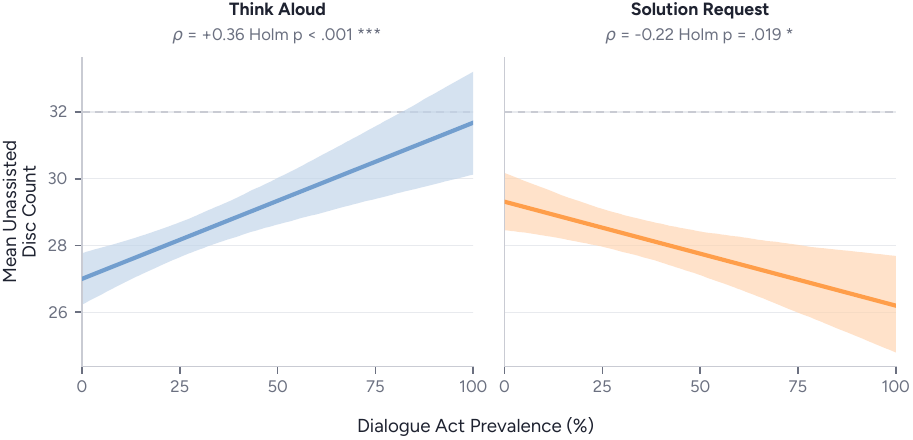}
    \caption{Relationship between Dialogue Act Prevalence and unassisted performance in the Othello experiments, $n = 111$. Transfer is the participant's own final disc count averaged over the two unassisted puzzles (higher is better). Each $8 \times 8$ board holds 64 discs, so 32 is an even board. Lines are least-squares fits and bands are 95\% percentile bootstrap intervals over participants; $\rho$ values are Spearman correlations with Holm-corrected significance across the two focal acts. The dashed gray line marks 32 discs, an even board.}
    \Description{Two side-by-side line charts sharing a vertical axis. The horizontal axis of each is Dialogue Act Prevalence from 0 to 100 percent; the vertical axis is the participant's mean final disc count across the two unassisted puzzles, running from about 25 to 33 discs, with a dashed reference line at 32 discs, an even board on the 64-square grid. The left chart is titled Think Aloud: its fitted line rises from 27.0 discs at 0 percent prevalence to 31.7 discs at 100 percent, with Spearman rho plus 0.36 and a Holm-corrected p below .001. The right chart is titled Solution Request: its line falls from 29.3 discs to 26.2, with rho minus 0.22 and p equal to .019. Each line carries a shaded 95 percent bootstrap confidence band that is narrowest in the middle of the range and widens toward both ends. 111 participants.}
    \label{fig:othello_dialogue}
\end{figure}

The Othello study reproduced the central behavioral pattern observed in Connect Four: participants who more frequently articulated their own reasoning performed better on subsequent unassisted puzzles, whereas participants who more frequently requested direct solutions performed worse. We recruited through the same channels as the Connect Four experiments (n=111 for the assisted/Standard-AI condition, n=100 for the baseline). We calculated Dialogue Act Prevalence values and their Spearman correlations with \textbf{the averaged disc count} (Figure \ref{fig:othello_dialogue}), focusing on \actTA{} and \actSR{} only, as they were the dialogue acts that most separated Solvers and Strugglers in the Connect Four trials. 

Both relationships replicated in Othello. \actTA{} and \actSR{} were the two focal behaviors identified in the discovery study, so we treated them as a prespecified family of two tests and corrected across them with the Holm procedure. Replicating the Connect Four result, participants who used \actTA{} more frequently during the AI interaction performed better in the transfer puzzles ($\rho = +0.36$, bootstrapped 95\% CI $[+0.20, +0.50]$, raw $p < .001$, Holm-adjusted $p < .001$; $n = 111$). \actSR{}'s DAP showed a negative correlation ($\rho = -0.22$, bootstrapped 95\% CI $[-0.40, -0.03]$, raw $p = .019$, Holm-adjusted $p = .019$). This reinforces our earlier conclusion: participants who \textbf{verbalize their reasoning} during the AI-assisted round are more likely to succeed on a subsequent, strategically related puzzle without assistance, while participants who \textbf{frequently ask for direct solutions} tend to perform worse when they solve the puzzles unassisted. Appendix~\ref{app:othello_robustness} reports a regression robustness check and rules out the reading that \actTA{} prevalence is simply a proxy for writing more. The same appendix adds the pre-assistance skill measures used in Connect Four. They are unrelated to \actTA{} DAP and to transfer, and the \actTA{} coefficient holds when they are included ($\beta = +0.352$ SD, $p < .001$, $n = 110$; Table~\ref{tab:prior_skill_othello}). These are associations between how participants used the assistant and how they later performed. Neither study manipulated conversational behavior.

\subsection{\texttt{Solution Request} is Associated with Stronger \textit{Assisted} Performance}
\actTA{} and \actSR{} behaved in opposite ways with respect to the \textbf{assisted} round (Table~\ref{tab:dissociation}). 

\noindent\actSR{} correlated strongly and positively with the assisted disc count ($\rho = +0.375$, $p < .001$) and negatively with transfer; \actTA{} showed the reverse, being unrelated to assisted performance ($\rho = -0.052$, $p = .60$) while carrying the unassisted performance association.

\begin{table}[t]
    \centering
    \small
    \begin{tabular}{lr@{\,}lr@{\,}l}
    \toprule
     & \multicolumn{2}{c}{\textbf{Assisted Discs}} & \multicolumn{2}{c}{\textbf{Unassisted Discs}} \\
     & \multicolumn{2}{c}{\small($n=101$)} & \multicolumn{2}{c}{\small($n=111$)} \\
    \cmidrule(lr){2-3} \cmidrule(lr){4-5}
    \textbf{Participant act} & \multicolumn{2}{c}{$\rho$} & \multicolumn{2}{c}{$\rho$} \\
    \midrule
      \actTA{}                & $-0.052$ &     & $+0.358$ & *** \\
      \actSR{}           & $+0.375$ & *** & $-0.222$ & *   \\
    \bottomrule
    \end{tabular}
    \caption{Dialogue Act Prevalence and the two outcomes, Spearman correlations, Othello
    participants. Both outcomes are the participant's own final disc count: on $O_{A}$, and averaged over $O_{U1}$ and $O_{U2}$. The assisted column requires that the assisted round be completed rather than stopped by the timer, whereas the transfer column imposes no such restriction.
    $^{*}p<.05$, $^{***}p<.001$.}
    \label{tab:dissociation}
\end{table}

\subsection{Assistant-side Behavior Links to Unassisted Performance}

The participant-side relationships had a corresponding assistant-side pattern: the kinds of responses produced by the AI themselves tracked subsequent unassisted performance.
To systematically analyze assistant-side behaviors, we utilized a modified version of the tutor-side taxonomy from \citeauthor{d2010mining} \cite{d2010mining} to code assistant utterances, using the same LLM judge panel as for user utterances. We record the resulting taxonomy in Table~\ref{tab:othello-assistant-taxonomy}. The panel models agree strongly on \actPCA{} (Fleiss' $\kappa = 0.82$ in Othello) and moderately on \actBR{} ($0.70$). Several acts that we do not analyze are coded less reliably; the feedback and paraphrase acts fall between $0.1$ and $0.6$ (Appendix~\ref{app:reliability}). Against a human coder on 50 held-out Othello replies, agreement was $\kappa = 0.81$ for \actPCA{} and $0.51$ for \actBR{}. Most \actBR{} disagreements fell on replies on which the panel's own models were split (Appendix~\ref{sec:taxonomy-human}), and we read the \actBR{} results with more caution than those for \actPCA{}.

\begin{table*}[t]
\centering
\small
\begin{tabularx}{\textwidth}{@{}l c X X@{}}
\toprule
Assistant act & \% of turns & Definition & Example from our Othello transcripts \\
\midrule
\actBR{} & 39 &
States or restates the current position without evaluating it, including e.g., occupied squares, legal moves, disc
counts. &
\textit{``Looking at the current board, b7 is occupied by a White disc---it's not a
legal Black move now. [\ldots] Your legal moves now are g2, b6, a7, and c8.''} \\
\addlinespace
\actPCA{} & 41 &
Names the move the assistant judges best. &
\textit{``I'd play a7.'' ``c8''.} \\
\addlinespace
\actGP{} & 60 &
States a strategic rule that holds beyond the position at hand. &
\textit{``a corner is almost always good, even early, because it can never be
flipped back.''} \\
\addlinespace
\actPrompt{} & 31 &
Asks the participant a question that directs attention to a feature of the board,
leaving the conclusion to them. &
\textit{``[\ldots] b8 or c8 are next to the empty corner a8 [\ldots]---could
they let White take a8 soon?''} \\
\addlinespace
\actHint{} & 26 &
Points toward the reasoning required without naming a move or completing the
argument. &
\textit{``Take a moment to compare the legal moves by corner access[\ldots]''} \\
\bottomrule
\end{tabularx}
\caption{The five assistant dialogue acts retained after filtering, with definitions and
verbatim examples drawn from participant-assistant transcripts. Percentages are the share
of the 616 assistant turns carrying each act. Acts are clause-level, so an assistant turn
typically carries several and the column does not sum to 100.}
\label{tab:othello-assistant-taxonomy}
\end{table*}

Assistant-side DAP results mirrored the participant-side relationships: \actBR{}, where the assistant explains and analyzes the current game state, correlated positively with unassisted performance ($\rho = +0.227$, $p = .017$, $n = 111$), while \actPCA{}, where the assistant supplies the optimal moves directly, exhibited a negative correlation ($\rho = -0.243$, $p = .010$). These assistant-side correlations were not prespecified and are reported as exploratory; we do not correct them across the five assistant acts. To further quantify this, we computed odds ratios that show how specific participant dialogue acts elicited different assistant dialogue acts (Table \ref{tab:othello-assistant-acts}).

\begin{table}[t]
\centering

\begin{tabular}{l c c c}
\toprule
 & \multicolumn{3}{c}{Participant act} \\
\cmidrule(l){2-4}
Assistant act & \actTA{} & \actSR{} & \actKDQ{} \\
\midrule
\actBR{}           & 1.20 [0.79, 1.81]  & 0.26 [0.16, 0.42]$^{***}$   & 1.97 [1.22, 3.16]$^{**}$  \\
\actPCA{} & 0.44 [0.29, 0.67]$^{***}$ & 12.95 [7.40, 22.66]$^{***}$ & 0.06 [0.02, 0.17]$^{***}$ \\
\actGP{}      & 0.96 [0.66, 1.38]  & 0.49 [0.33, 0.73]$^{***}$   & 4.21 [2.40, 7.37]$^{***}$ \\
\actPrompt{}                 & 2.31 [1.56, 3.44]$^{***}$ & 0.16 [0.09, 0.28]$^{***}$ & 0.49 [0.28, 0.86]$^{*}$   \\
\actHint{}                   & 1.85 [1.17, 2.93]$^{**}$  & 0.16 [0.08, 0.33]$^{***}$ & 0.48 [0.24, 0.95]$^{*}$   \\
\bottomrule
\end{tabular}

\caption{Adjusted odds ratios with $95\%$ confidence intervals for each assistant act in the Othello study, from a logistic model of the assistant act on three participant acts jointly to account for co-occurrence within a turn ($\textrm{assistant act} \sim \textrm{\actTA{}} + \textrm{\actSR{}} + \textrm{\actKDQ{}}$). Participant turns are nested within participants, so standard errors are clustered on participant ($616$ turns, $111$ participants). Values above 1 indicate the assistant act is more likely; below 1, less likely. $^{*}p<.05$, $^{**}p<.01$, $^{***}p<.001$}
\label{tab:othello-assistant-acts}
\end{table}

Across participant acts, \actBR{} was never the most likely assistant response. \actKDQ{} significantly increased the likelihood of \actBR{}, but most often elicited \actGP{}. Upon further inspection, we additionally identified \textbf{temporal orientation} as a factor: retrospective (e.g., ``was that a bad move?'') versus prospective (e.g., ``is a1 the best move now?'') questions. Retrospective utterances were significantly more likely to elicit \actBR{} ($OR = 3.13, p < 10^{-6}$). Orientation is assigned by a deterministic rule over part-of-speech features (past tense and perfect aspect for retrospective; modals, future cues and bare move coordinates for prospective), so it is reproducible exactly. Against a human coder on 100 Othello turns, the rule agrees well on retrospective orientation ($\kappa = 0.83$), the distinction this result rests on, and less on prospective ($0.51$) and current-state ($0.22$) orientation.

Through our analysis, the behavioral profile that accompanies stronger unassisted performance begins to emerge: someone who articulates their own reasoning, engages in reflection and retrospection, and resists the urge to ask for the best moves directly. These observational results do not show that participant behavior \textit{causes} assistant responses. We ran controlled counterfactual replays (Appendix~\ref{app:controlled_replay}), rewriting real participant utterances along one conversational dimension while keeping core content, conversation history, board state, and the assistant fixed. Assistant behavior shifted as the correlations predicted: making an utterance a direct next-move request increased direct answers, while reframing it as a retrospective question about the previous move increased board-state descriptions. These effects were consistent across all four assistant models we examined. Our causal claim is narrow: such conversational choices can change the assistant's immediate response, not participants' later unassisted performance.


\subsection{AI Assistance Improves Assisted Performance but Does Not Reliably Improve Transfer}
\label{sec:assistance-vs-transfer}

Compared with baseline participants ($n=100$), assisted (Standard-AI) participants ($n=111$) finished the \textit{assisted} round holding significantly more discs ($27.26$ against $21.52$ of $64$; Mann-Whitney $U = 8186$, $p < .001$) and rated their own performance higher on the NASA-TLX ($3.14$ against $2.01$; $U = 7439$, $p < .001$). AI assistance significantly improved participant performance in the assisted task, consistent with previous work on AI and productivity \cite{al2024enhancing,przegalinska2025collaborative,noy2023experimental}.

In the unassisted transfer rounds ($O_{U1}$ and $O_{U2}$), the difference between assisted and baseline performance became weaker. Standard-AI participants held slightly more discs, but the difference was not significant ($+0.25$ discs per game, 95\% CI $[-0.58, +1.09]$, $d = 0.08$, $p = .55$).

\begin{figure}
    \centering
    \includegraphics[width=0.8\linewidth]{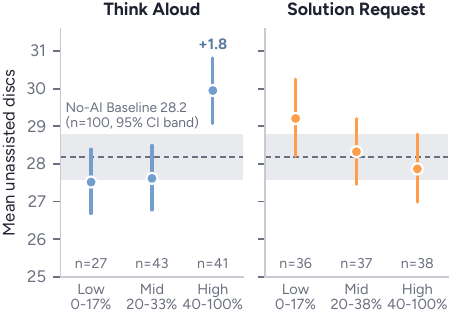}
    \caption{Unassisted transfer of Standard-AI participants, grouped by the share of their turns carrying each act, against the no-AI Baseline ($n = 100$; dashed line, with its 95\% bootstrap CI shaded). Points are group means of the disc count averaged over $O_{U1}$ and $O_{U2}$; whiskers are 95\% bootstrap CIs. The label marks the one group whose difference from the Baseline survives Holm correction over the six comparisons.}
    \Description{Two side-by-side dot-and-whisker panels sharing a vertical axis of mean unassisted disc count, from 25 to about 31.5. A dashed horizontal line at 28.2 marks the no-AI Baseline mean, with a grey band from about 27.6 to 28.8 for its 95 percent interval. Left panel, Think Aloud: three groups labelled Low (0 to 17 percent of turns, n = 27, mean 27.5), Mid (20 to 33 percent, n = 43, mean 27.6) and High (40 to 100 percent, n = 41, mean 30.0). Low and Mid sit just below the Baseline with whiskers crossing it; High sits well above it with a whisker from about 29.0 to 30.8, and is labelled +1.8. Right panel, Solution Request: Low (0 to 17 percent, n = 36, mean 29.2), Mid (20 to 38 percent, n = 37, mean 28.3) and High (40 to 100 percent, n = 38, mean 27.9); all three whiskers overlap the Baseline band.}
    \label{fig:baseline_terciles}
\end{figure}

Further leveraging participant baseline performance to interpret participant behavioral patterns, we split Standard-AI participants by the prevalence of our acts of interest and compared each group with the Baseline (Figure~\ref{fig:baseline_terciles}). We divide participants into three groups based on prevalence for each act: below 20\%, 20-40\%, and at least 40\% of turns. Only one group differed significantly from the Baseline. Participants who thought aloud on at least 40\% of their turns held $+1.78$ more discs per unassisted game (95\% bootstrap CI $[+0.69, +2.84]$, $d = 0.59$). In this sample, the \actTA{} association reflects a gain over unassisted play among those who articulated their reasoning most, rather than a loss among those who did not.

Perceived performance, however, behaved differently. For the NASA-TLX administered after the unassisted round, \textbf{assisted participants rated their task success significantly higher than baseline} (2.95 vs.\ 2.51, $d = 0.32$, $p = .007$), but none of the other items differed significantly. Perceived success correlated moderately with actual disc count ($r = 0.41$, $p < .001$), both among Standard-AI participants ($r = 0.41$) and Baseline participants ($r = 0.43$).

In sum, standard AI assistance yielded substantial performance benefits when available, but we did not observe a reliable advantage in unassisted rounds. \textbf{Assisted performance was, in fact, negatively correlated with unassisted performance} ($\rho = -0.307$, $p = .002$, $n = 97$). This result motivates us to study differences in \textit{how} users engage with AI, as providing access to AI alone does not reliably produce transfer.

\subsection{Additional Validation Under an External Taxonomy}
\label{sec:turnstyle}

Our taxonomy was adopted from \citeauthor{d2010mining} and then enriched with task-specific dialogue acts, which raises a question about the generalizability of our behavioral coding. We therefore re-described the same conversations under TurnStyle, a taxonomy built by other authors, for other tasks, without reference to our data \cite{awasthi2026turnstyle}.

\paragraph{Annotation} We labeled every participant turn in the Othello study ($616$ turns across $111$ conversations) with a TurnStyle subcategory, then mapped subcategories to the six top-level categories, following the original procedure. Labels were produced with TurnStyle's published annotation prompts, so the categories are applied as their authors defined them; we did not re-validate them against human labels on our data. We treat this section as a convergent check, not as an independent measurement.

\paragraph{The same behavior correlates with assisted and transfer outcomes in opposite directions}
Assisted performance on $O_{A}$ and unassisted performance on $O_{U1}$ and $O_{U2}$ were themselves negatively related ($\rho = -0.307$, $p = .002$, $n = 97$). Against these two outcomes, the same behavioral measures pointed in opposite directions (Table~\ref{tab:turnstyle}).

\begin{table}[t]
\centering
\small
\begin{tabular}{@{}l r r@{}}
\toprule
TurnStyle category & $O_{A}$ & $O_{U1, U2}$ \\
\midrule
Task management        & $+0.460^{***}$ & $-0.280^{**}$ \\
Information request    & $-0.430^{***}$ & $-0.037$ \\
Reasoned agreement     & $-0.311^{**}$  & $+0.190^{*}$ \\
Clarification request  & $+0.150$       & $+0.137$ \\
Off topic              & $-0.190$       & $-0.074$ \\
Critical evaluation    & $+0.145$       & $+0.101$ \\
\bottomrule
\end{tabular}
\caption{Spearman correlations between each TurnStyle category's share of a participant's
turns and each outcome; both outcomes are the participant's own final disc count
($n=101$ assisted, $n=111$ unassisted). This re-description under an external taxonomy is exploratory:
the six categories are tested against two outcomes without correction. $^{***}p<.001$, $^{**}p<.01$, $^{*}p<.05$.}
\label{tab:turnstyle}
\end{table}

Task management, in our case mapping to asking the assistant to supply the move, was the strongest single covariate of doing well when assisted, and a significant negative covariate of unassisted performance. This is a replication of our \actSR{} result. Reasoned agreement reversed direction, negative on the assisted round and positive on transfer, while information request was negatively correlated with assisted performance and unrelated to transfer. 

Scored against the assisted outcome, TurnStyle's inquiry-loop finding replicated more strongly in our data: participants whose conversation contained an Information-request loop won the assisted round far less often ($OR = 0.12$, $95\%$ CI $[0.01, 0.91]$, $p = .020$, $n = 101$). However, other prior findings pertaining more to open-ended collaboration did not appear here. We attribute these to task structure rather than to either framework, as we focused on puzzles with fixed solutions.

\section{Case Study: Qualitative Look at Exemplar Conversations}
\label{sec:cases}

In this section, we present four case studies of anonymized participant transcriptions to characterize what the aggregate measures do not capture.  These participants were selected as extreme instances of the outcome patterns the aggregate results describe (Table~\ref{tab:adopter-profiles}): two who succeeded with assistance and failed without it, one who succeeded at both, and one who failed with assistance yet transferred well. Complete transcripts appear in Appendix~\ref{app:conversations}.

\paragraph{Game Play Binary Labels}
All Othello participants who completed the assisted ($O_{A}$) and unassisted tasks ($O_{U1}$ and $O_{U2}$), $n=97$, were assigned binary labels to split them into successful or unsuccessful gameplay groups. Based on a participant's performance on the assisted and unassisted tasks, they were labeled as 1-1, 0-0, 1-0, or 0-1 (Table~\ref{tab:quadrants}). A participant labeled ``1-0'' was ``successful'' in the assisted but ``unsuccessful'' in the unassisted tasks.  

\begin{figure}
    \centering
    \includegraphics[width=\linewidth]{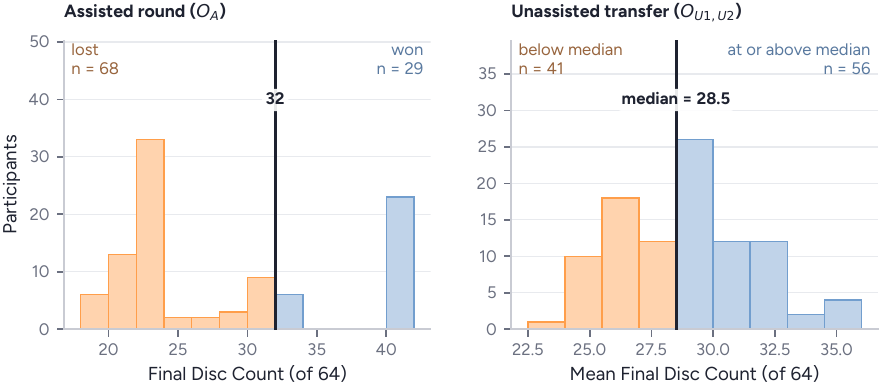}
    \caption{The distribution of assisted and unassisted task performance among the $n=97$ Othello participants who finished every round by play. Performance is the participant's own final disc count out of 64, averaged over the two unassisted puzzles on the right. The thresholds used to separate ``successful'' and ``unsuccessful'' participants are denoted with vertical lines.}
    \Description{Two histograms of participant performance, both measured as the participant's own final disc count, for the 97 Othello participants who finished every round by play. The left histogram, titled ``Assisted round'', is strongly bimodal: a large cluster of bars sits between about 18 and 31 discs and a second, separate cluster sits near 41 discs, with a solid vertical line at 32 discs marking an even board. 29 participants fall above that line and 68 at or below it. The right histogram, titled ``Unassisted transfer'', shows the mean disc count across the two unassisted puzzles; it is unimodal and mildly right-skewed, spanning roughly 23 to 36 discs, with a solid vertical line at the median of 28.5 discs. 56 participants fall at or above the median and 41 below it. Bars left of each threshold are orange and bars right of it are blue.}
    \label{fig:scores_dist}
\end{figure}

The thresholds splitting participant ``success'' were determined based on the distribution of scores across all participants (Figure \ref{fig:scores_dist}). In the assisted task, participant scores were bimodal and split at 32 discs, an even board; meanwhile, in the unassisted tasks, scores (the mean of $O_{U1}$ and $O_{U2}$) were unimodal and mildly right-skewed (skew $= 0.52$; Hartigan's dip test on the dithered scale, $p = .32$) and therefore split along the median (threshold $= 28.5$ discs). 
\label{sec:quadrants}





\begin{table}[t]
    \centering
    \small
    \begin{tabular}{lcccc}
    \toprule
      \textbf{Assisted Label} & \textbf{Unassisted Label} & \textbf{$n$} & \textbf{\actTA{}} & \textbf{\actSR{}} \\
    \midrule
      1  & 1 & 11 & 42\% & 34\% \\
      1  & 0 & 18 & 17\%& 46\% \\
      0 & 1 & 45 & 33\% & 21\% \\
      0 & 0 & 23 & 26\% & 26\% \\
    \bottomrule
    \end{tabular}
    \caption{Participants partitioned by assisted and transfer outcome, with mean \actTA{} and \actSR{} prevalence per cell. Among participants who were successful in the assisted round, those who were also successful in the unassisted round used more thinking aloud than those who were unsuccessful (42\% vs 17\%). Similarly, among participants who were unsuccessful in the assisted round, those who were successful in the unassisted round also used more thinking aloud (33\% vs 26\%). Respectively, solution request usage was flipped.}
    \label{tab:quadrants}
\end{table}

\begin{table*}[t]
  \centering
  \small
  \begin{tabularx}{\textwidth}{@{}lXXXX@{}}
    \toprule
      & \textbf{P1} & \textbf{P2} & \textbf{P3} & \textbf{P4} \\
    \midrule
      Label & 1-0 & 1-1 & 1-0 & 0-1 \\
      Assisted discs & $33$ & $41$ & $41$ & $19$ \\
      Unassisted discs (mean) & $24.0$ & $36.0$ & $26.0$ & $36.0$ \\
      \quad Puzzle \#2 / \#3 & $26$ / $22$ & $38$ / $34$ & $26$ / $26$ & $38$ / $34$ \\
    \addlinespace
      Turns & 5 & 9 & 10 & 6 \\
      Think Aloud & 0.00 & 0.78 & 0.30 & 0.67 \\
      Solution Request & 1.00 & 0.11 & 0.50 & 0.00 \\
    \addlinespace
      Self-rated success & 6 $\rightarrow$ 1 & 7 $\rightarrow$ 7 & 7 $\rightarrow$ 2 & 2 $\rightarrow$ 5 \\
      Effort & 1 $\rightarrow$ 4 & 2 $\rightarrow$ 3 & 4 $\rightarrow$ 6 & 4 $\rightarrow$ 5 \\
      Frustration & 1 $\rightarrow$ 5 & 1 $\rightarrow$ 1 & 2 $\rightarrow$ 2 & 5 $\rightarrow$ 5 \\
      Mental demand & 1 $\rightarrow$ 4 & 3 $\rightarrow$ 3 & 5 $\rightarrow$ 6 & 5 $\rightarrow$ 5 \\
    \addlinespace
      Matched my intent & Strongly agree & Strongly agree & Strongly agree & Agree \\
      Easy to use & Strongly agree & Neutral & Strongly agree & Neutral \\
    \addlinespace
      Age & 69 & 35 & 32 & 34 \\
      Education & postgrad & postgrad & bachelors & postgrad \\
      GenAI use & daily & about monthly & several times day & daily \\
      Self-rated skill & 1 & 1 & dont know & 1 \\
    \bottomrule
  \end{tabularx}
  \caption{The four exemplar adopters. Scores are the participant's own final disc count out of 64; transfer is the mean over the two unassisted puzzles. Think Aloud and Solution Request are prevalences over the participant's turns. NASA-TLX items run 1--7, and the \emph{performance} item asks how successful the participant felt, so a higher value denotes greater self-assessed success; arrows run from the assisted round to the final unassisted round. Self-rated skill is the pre-study item, on which 4 and 5 were screened out.}
  \label{tab:adopter-profiles}
\end{table*}

\newtcolorbox{excerpt}[1][]{colback=black!3, colframe=black!30, boxrule=0.4pt,
  left=4pt, right=4pt, top=3pt, bottom=3pt, arc=1pt, fonttitle=\bfseries\small,
  coltitle=black, colbacktitle=black!8, #1}

\newcommand{\excerptPone}{%
\begin{excerpt}[title={P1 (p258): assisted $33$ discs, transfer $24.0$}]
\small
\textbf{P (SR):} \textit{``where?!''}\\[2pt]
\textbf{AI:} Play \textbf{b7}. It's a quiet edge move that avoids handing White an easy corner right away, and it keeps Black's endgame chances best.\\[2pt]
\textbf{P (SR):} \textit{``now?''}\\[2pt]
\textbf{AI:} Play \textbf{c8}. It's the best endgame move here: it keeps pressure on the bottom edge without giving White an immediate easy corner swing.
\end{excerpt}}

\newcommand{\excerptPtwo}{%
\begin{excerpt}[title={P2 (p322): assisted $41$ discs, transfer $36.0$}]
\small
\textbf{P (TA):} \textit{``white cannot take h1 if i make that move''}\\[2pt]
\textbf{AI:} Let's verify that line: after \textbf{Black g2}, the diagonal \textbf{h1--g2--f3} would be: h1 empty g2 Black f3 White So \textbf{White \textit{can} play h1} next, flipping g2. That's the danger: g2 wins pieces now, but may hand over a permanent corner.\\[2pt]
\textbf{P (TA, CA):} \textit{``oh i see i didnt realise diagonal counts''}
\end{excerpt}}

\newcommand{\excerptPthree}{%
\begin{excerpt}[title={P3 (p161): assisted $41$ discs, transfer $26.0$}]
\small
\textbf{P (SR):} \textit{``im looking for the best possible move to maximise success. Not just a good move, but statistically the best possible move''}\\[2pt]
\textbf{AI:} Best move for Black: \textbf{b7}. Even though b7 is next to the empty corner \textbf{a8}, in this exact endgame it gives Black the best outcome. \textit{[\ldots]}\\[2pt]
\textbf{P (CGQ, MC):} \textit{``have you worked out any other possible outcomes and you are certain this is a strong choice? you are free to take the time needed to run the numbers''}\\[2pt]
\textbf{AI:} Yes --- \textbf{b7 is the best move I'd trust here}. This isn't a ``flip the most now'' choice; it's an endgame choice. The important points: \textbf{b7 flips the whole 7th row across to g7}, giving Black a strong edge structure. \textit{[\ldots]}
\end{excerpt}}

\newcommand{\excerptPfour}{%
\begin{excerpt}[title={P4 (p786): assisted $19$ discs, transfer $36.0$}]
\small
\textbf{P (TA):} \textit{``now b6 just lets white retake that diagonal. even if a7 gives white b8 so the 8 row that might be better''}\\[2pt]
\textbf{AI:} Yes --- that's the right trade-off to notice. Small correction: after \textbf{a7}, White's danger is taking the \textbf{corner a8}, which can flip along row 8 through \textbf{b8--e8}. \textit{[\ldots]}\\[2pt]
\textbf{P (TA, CGQ):} \textit{``so what i want to minimise is whites corners? more than justc white edges?''}\\[2pt]
\textbf{AI:} Yes --- \textbf{White corners matter much more than White edges}. \textbf{Corners are permanent}: once White gets a corner, it can't be flipped back. A White corner often makes the nearby \textbf{edge discs stable too}. \textit{[\ldots]}
\end{excerpt}}

\subsection{P1: Efficient Assistant Consultation without Intuition Building; Label: 1-0}

P1 is a typical instance of effectively leveraging the AI assistant by directly probing it for answers, without any observable effort to build intuitions about Othello. This is the pattern that motivates concern about AI assistance eroding the capability it is meant to boost: the assistant answers the question that was asked, the participant acts on the answer, and nothing in the exchange requires the participant to build an understanding of the board that would survive the assistant's removal.

\begin{quote}
    \textbf{P1:} \textit{``[W]here?!''}
\end{quote}

This participant's first utterance is clear \actSR{}, without any stated grounding on the board state nor reflection. Their subsequent four utterances are all \textit{``now?''}.

The assistant responses are similarly terse, starting with \actPCA{}, then often transitioning to \actGP{}. 

\begin{quote}
    \textbf{Assistant:} \textit{``Play \textbf{b7}.}

    \textit{``It's a quiet edge move that avoids handing White an easy corner right away, and it keeps Black's endgame chances best.''}
\end{quote}

The participant did not take time to peruse the provided rationales, as the entire conversation lasted 68 seconds: sufficient time for making moves, likely insufficient to digest and internalize stated principles.

P1 narrowly won the assisted task ($33$ of $64$ discs), but failed when unassisted, averaging $24.0$ discs against a median of $28.5$. This shows that pure delegation may achieve high performance, yet unassisted performance may suffer. Their \actSR{} prevalence was 100\%; in other words, every prompt they used was a direct solution request.

\subsection{P2: Stating Reasoning and Being Corrected; Label 1-1}

P2's conversation consists of a sequence of move proposals with attached reasoning, each of which the assistant can engage with because the reasoning is visible. 

\begin{quote}
    \textbf{P2:} \textit{``I am thinking of going in 2g because it looks like it will gain me the most pieces[. W]hat do you think[?]''}

    \textbf{Assistant:} \textit{``Be careful: g2 is the X-square next to the empty corner h1. If you play g2, ask yourself: [\ldots] In endgames, don't count just this move's flips---look at White's reply, corners, and who gets the last empty squares. [\ldots]''}
\end{quote}

The participant's utterance is a clear case of \actTA{}. Throughout the conversation, P2 exhibited two misconceptions that the assistant corrected. Such corrections would not have been possible if P2 had not stated their reasoning in the first place. 

\begin{quote}
    \textbf{P2:} \textit{``[W]hite cannot take h1 if [I] make that move''}

    \textbf{Assistant:} \textit{``Let's verify that line: after Black g2, the diagonal h1--g2--f3 would be: h1 empty, g2 Black, f3 White. So White can play h1 next, flipping g2. That's the danger: g2 wins pieces now, but may hand over a permanent corner.''}

    \textbf{P2:} \textit{``[Oh, I] see [I didn't realize] diagonal counts''}
\end{quote}

As a result of the correction, P2 could have experienced an update to their internal model of the game. This offers an alternative view as to why \actTA{} is germane to transfer, as it involves stating such rationales explicitly. 

P2 won the assisted round ($41$ of $64$ discs) and was tied for the highest transfer score in the study ($36.0$ discs, shared with three other participants). Their \actTA{} prevalence was 78\%, and their \actSR{} prevalence was 11\%.

\subsection{P3: Scrutinizing the Assistant, Not the Board; Label: 1-0}

P3 demonstrates high proficiency in using AI assistance, requiring careful consideration by the assistant. However, across ten turns, the scrutiny is often directed at the assistant's confidence rather than at the board. 

\begin{quote}
    \textbf{P3:} \textit{``[H]ave you worked out any other possible outcomes and you are certain this is a strong choice? [Y]ou are free to take the time needed to run the numbers.''}

    \textbf{P3:} \textit{``[O]kay [I] played B7 - [what's] the next move. We have 3 more turns to finish my position, [so let's plan] to have 3 more [consecutive] turns''}
    
\end{quote} 

The second utterance qualifies as a \actTA{} utterance because the intention of planning is demonstrated. However, the verbalized rationale is still directed at the interaction with the assistant, instead of with respect to the board. 

P3 succeeded in the assisted task ($41$ discs) but failed in the unassisted task ($26.0$ discs, below the $28.5$ median), showing poor transfer skills. Their transcriptions show that 30\% of their speech acts were \actTA{}, meanwhile their \texttt{\actSR{}} rate was at 50\%.

\subsection{P4: Extracting a Principle from a Loss; Label: 0-1}

P4 reasons extensively about Othello's general principles, asking multiple questions about gameplay heuristics to facilitate unassisted performance retention. 

\begin{quote}
    \textbf{P4:} \textit{``[Am I] right in thinking that [I] should choose edge squares[,] not middle squares[,] most of the time?''}

    \textbf{P4:} \textit{``[So] what [I] want to minimize is [white's] corners? [M]ore than [just] white edges?''}
\end{quote}

By asking about principles, P4 was able to leave the round retaining generalizable Othello principles. During the assisted round, P4 additionally reasoned through a trade-off on the board and then abstracted from it. Despite not performing well during the assisted round, P4 performed better without the assistant.

In the post-study surveys, participants had the option to provide free-text feedback to the assistant. Therefore, we could observe how users felt about the interaction. Paradoxically for P4, they rated the assistant lowest of our four cases and described the assistant as not helpful, even when they achieved the best unassisted performance. P4 articulated the mismatch precisely: \textit{``what [I] want from it is (1) which move is best and (2) what general strategy should [I] use, whereas it tended to just tell me to think about [minimizing opponents'] options - not helpful.''}

P4 lost the assisted round badly ($19$ of $64$ discs) but was tied for the highest transfer score in the study ($36.0$ discs). Their \actTA{} rate was 67\%, and they never once requested a solution.

\section{Think-Aloud Subtypes Elicit Assistant Acts Differently}
\label{sec:prompt-types}

Our case study highlights distinctions that our taxonomy cannot express: P2 engages deeply with the game state, and P3 criticizes the assistant; both instances are categorized as \actTA{}. While we could propose constructs based on the observed differences in the chosen dialogues, we observed that participants with successful transfer did not always follow the same conversational recipe. As a result, we decided to leverage a more quantitative, unsupervised approach to cluster \actTA{}.

We leveraged ConvoKit's \texttt{PromptTypeWrapper} tool to cluster all 193 \actTA{} utterances \cite{zhang2017asking, chang2020convokit}. The tool groups conversational utterances by the type of reply they elicit in a deterministic manner, making it the ideal instrument for our purpose. Out of 193 utterances, 139 received a type assignment, as the remainder carried no usable dependency arcs.

The unsupervised tool separated \actTA{} into interpretable clusters. We report $k=4$ here for readability; robustness across $k$ is addressed below. The largest cluster ($n=44$) consists of turns that ask why something on the board happened or what would follow from a move: \textit{``I chose a 7 but it was an illegal move, why?''}, \textit{``I played c8 but [W]hite is in the a8 corner now?''}, \textit{``[W]hat if I did g2? [W]ould [W]hite have a legal move?''} We refer to these as \textbf{diagnostic} turns. A second cluster of comparable size ($n=39$) names a candidate square and asks the assistant to provide a selection: \textit{``[I'm] thinking c8, is that any good?''}, \textit{``I'm considering G2. Does this work?''}, \textit{``Is a7 good? [T]hat's my initial thought''}. We refer to these as \textbf{verification} turns. A third cluster ($n=41$) reports board state or intention without posing a question at all (\textit{``[I] can't choose between b6 and a7''}, \textit{``[I] think my next move should be a7''}), which we call \textbf{narration}, and a small fourth cluster ($n=15$) delivers a verdict on a move, sometimes with a reason attached (\textit{``g2 looks good because [W]hite can't play the corner next''}), which we call \textbf{evaluative}. 

These subtypes differed sharply in what they drew from the assistant (Table~\ref{tab:ta-subtypes}). Verification turns were followed by a \actPCA{} at $OR = 6.71$ ($95\%$ CI $[3.55, 12.65]$, $p < .001$), similar in magnitude to how likely a \actPCA{} follows a \actSR{}, while diagnostic turns suppressed it at $OR = 0.23$ ($95\%$ CI $[0.07, 0.69]$, $p = .009$). One flavor of thinking aloud, therefore, pulls much the same response from the assistant as a \actSR{} does, which the act taxonomy alone would not reveal. This result does not depend on the number of clusters, as refitting at different $k$ produced \actTA{} subtypes capable of inducing \actPCA{} from the assistant compared to other subtypes.



We are careful about what this supports. The claim we make is about subtypes of \actTA{} behaving differently: they vary systematically in whether they draw a direct answer from the assistant. Although verification-type \actTA{} is significantly more likely to elicit \actPCA{}, it \textit{does not itself predict transfer}. This is what we would expect given how few reasoning turns most participants produce, so it cannot rule out a role for subtype either. We therefore claim only the elicitation result, not that the type of \actTA{} utterance mediates unassisted performance.

\begin{table}[t]
    \centering
    \small
    \begin{tabular}{lcr@{\,}lr@{\,}l}
    \toprule
     & & \multicolumn{2}{c}{\textbf{\actPCA{}}} & \multicolumn{2}{c}{\textbf{\actBR{}}} \\
    \cmidrule(lr){3-4} \cmidrule(lr){5-6}
    \textbf{\actTA{} subtype} & $n$ & \multicolumn{2}{c}{$OR$} & \multicolumn{2}{c}{$OR$} \\
    \midrule
      Verification & 39 &  6.71 & *** & 0.65 & \\
      Diagnostic   & 44 &  0.23 & **  & 0.96 & \\
      Evaluative   & 15 &  0.43 &     & 1.04 & \\
      Narration    & 41 &  1.32 &     & 1.26 & \\
      \addlinespace
      \multicolumn{6}{@{}l}{\textit{for comparison}} \\
      \quad \actSR{} & 175 & 18.34 & *** & 0.29 & *** \\
    \bottomrule
    \end{tabular}
    \caption{What each \actTA{} subtype elicits from the assistant, at $k=4$. Odds ratios (ORs) are turn level with standard errors clustered on participant; the reference
    class for each subtype is the remaining \actTA{} turns, and for \actSR{} it is all other participant turns. 
    $^{*}p<.05$, $^{**}p<.01$, $^{***}p<.001$.}
    \label{tab:ta-subtypes}
\end{table}

\section{Discussion}

Across our studies, we find that participants' performance without AI assistance tracks the distribution of reasoning workload between the user and the assistant. Interactions in which users articulate their reasoning correspond to better unassisted performance, and more instances of direct requests for the next best moves to poorer transfer. These participant-side behaviors drive assistant responses accordingly: elaborating thought processes elicits more tutor-like behavior, while soliciting solutions results in direct answers. Finally, qualitative and subtype analyses reveal substantial heterogeneity within the aggregate \actTA{} category. We discuss our findings in the context of cognitive engagement and delegation, the need for users to steer AI assistants towards pedagogical roles, and the limits of using coarser-grained dialogue taxonomies.

\subsection{Thinking Aloud as Cognitive Participation during AI Assistance}

\actTA{} tracks better unassisted performance in both Connect Four and Othello studies, while\\ \actSR{} is coupled with worse unassisted performance in Othello. These two modes of interaction with AI are instances of retaining vs. delegating cognitive work. \textbf{\actTA{} is \textit{consistent} with}, albeit broader than, \textbf{constructive engagement} in the ICAP framework \cite{chi2014icap} and \textbf{self-explanation} \cite{chi1994eliciting,bisra2018inducing}, because although \actTA{} features verbalized thought, not all of these utterances may constitute productive explanation. We find that the \textit{rate} of \actTA{} relates positively to unassisted performance, a quantity-based pattern that echoes \citeauthor{chi1989self}: ``Good'' students produce many explanations, and relate the principles of the materials, while ``Poor'' students do not generate sufficient self-explanations and fail to monitor their learning accurately \cite{chi1989self}. While the \textit{type} of self-explanations has also been shown to contribute to learning outcomes \cite{renkl1997learning}, this is yet to be strongly supported in our study through our subtype analyses. Our work differs from existing self-explanation literature: traditional self-explanations are directed toward students themselves, whereas we include a responsive interlocutor.


Meanwhile, \textbf{\actSR{} maps neatly to executive help-seeking} \cite{martin2021help,nelson1981help}. Existing research on help-seeking distinguishes between two primary modes of help-seeking: \textit{executive} help-seeking, where students expect the problem to be solved for them, and \textit{instrumental} help-seeking, where students request only the help required to complete the task on their own \cite{nelson1981help,karabenick2004perceived,karabenick2013help,martin2021help}.
Executive help-seeking is known to improve immediate progress while producing worse learning outcomes \cite{aleven2003help,karabenick2003seeking}; instrumental help-seeking, we hypothesize, is more likely to support future unassisted performance.
\actSR{}, the act of requesting the next optimal move directly, transfers the agency for game state interpretation and action selection to the assistant and is therefore consistent with the definition of executive help-seeking. As a result, the user may fail to learn skills required to produce good performance when AI assistance is no longer available. We emphasize that our results do not indicate that productive human-AI collaboration must avoid executive help-seeking. In our Connect Four study, Solvers still invoke \actSR{}, but their requests co-occur with \actTA{}. Likewise, \actSR{} appears in one of our strongest Othello case-study conversations, although \actTA{} dominates. In sum, these patterns are consistent with transfer being stronger when users remain involved in constructing and monitoring their own representation of the problem. \citeauthor{newell1972human} describe this representation as a \textit{problem space}: the states a solver can be in, the operators that move between them, and the tests that recognize a goal \cite{newell1972human}. In these terms, \actTA{} has users search the problem space themselves, proposing operators and evaluating the states they lead to, whereas \actSR{} hands operator selection to the assistant, so users can reach a good state without building the space that produced it. A new puzzle requires that space again, which may explain why \actSR{} accompanies better assisted performance but not better transfer.

Importantly, the user's ability to monitor and build mental representations is not necessarily visible in their immediate assisted performance, since \textbf{assisted performance does not always reliably indicate strong unassisted performance}. This is validated through the negative correlation in our Othello study, and further demonstrated in Section \ref{sec:turnstyle}, where we re-code our collected conversations using an external taxonomy developed to identify in-task human-LLM interaction dynamics. \citeauthor{awasthi2026turnstyle} find that concrete task management is beneficial to task success, which is affirmed through the positive correlation between \actSR{} and assisted performance \cite{awasthi2026turnstyle}. However, the same behaviors that improve in-task performance accompany worse transfer. This means that curricula targeting human-AI collaboration performance may overlook the retention of performance without AI assistance.

We recognize that our data does not rule out the possibility that \actTA{} is a \textit{symptom} of participants who already attend closely to the board and would be inclined to extract learnable principles from the game regardless. The skill measures we hold from before the assistant was available do not explain the association. They are uncorrelated with \actTA{}, and adding them leaves its coefficient intact in both studies (Section~\ref{sec:cf-prior-skill}, Appendix~\ref{app:othello_robustness}). But these measures barely predict unassisted performance even in our no-AI Baseline. A null adjustment on weak proxies does not settle the question, so we read every result in this paper as an association between conversational behavior and later performance, not as an effect of that behavior. To firmly demonstrate that articulating reasoning is a learnable user strategy that enables productive AI interactions to translate into retention, future empirical work should test whether interventions that increase the prevalence of \actTA{} also improve unassisted performance. We outline design guidelines for enabling such interactions in Section \ref{sec:design_imp}.

\subsection{Users Shape the Pedagogical Behavior of Less Constrained Assistants}

We observe a symmetry between participant-side and assistant-side behavioral patterns that track transfer. In our Othello trials, \actBR{} is positively correlated with transfer outcomes, while this relationship is negative for \actPCA{}. \actTA{} tends to evoke more tutor-like behaviors, whereas \actSR{} draws out direct answer provision. Users' conversational choices therefore actively shape subsequent assistance received. This relationship is further validated by our counterfactual rewriting experiments (Appendix \ref{app:controlled_replay}), which establish that minimally changing the form of a user's request can causally alter the assistant's response behavior. More broadly, these results suggest that an AI assistant can be steered toward explanation, prompting, or direct solution provision through users' conversational choices.

\subsection{\texttt{Think Aloud} Collapses Multiple Interaction Strategies}

Although \actTA{} correlates with stronger unassisted performance across our studies, our case study and subtype analyses indicate that the same label does not reflect a single interaction strategy. Prior work has similarly identified different types of self-explanations, ranging from principle-based explanations to elaboration of the problem situation \cite{renkl1997learning}. Indeed, we can separate \actTA{} utterances collected in our studies into interpretable clusters, and its different subtypes exhibit distinct behavior during human-AI interactions. 

With increasing granularity in our subtype analysis, we identify stable subtypes of \actTA{}: verification and diagnostic. 
Verification subtype turns often offer move candidates for evaluation, whereas diagnostic turns ask the assistant to reason about why an outcome occurred or what would follow from a move. These differences matter for the assistance elicited, since only verification turns have substantially higher odds of eliciting \actPCA{}. Therefore, two utterances can both satisfy our \actTA{} definition while allocating different parts of the reasoning process to the assistant.

We do not find that these subtypes themselves robustly predict transfer, so we do not interpret them as identifying intrinsically better or worse forms of \actTA{} in our setting. Rather, they show that the aggregate label collapses interactionally distinct behaviors and that frequency alone does not fully characterize the distribution of reasoning within the conversation.

\section{Design Implications}
\label{sec:design_imp}

Our study indicates that conversational behaviors during human-AI interaction are correlated with user performance when the AI is removed. The implications below are therefore hypotheses to test, not prescriptions. If \actTA{} marks users who are already engaged rather than making them so, nudging other users to verbalize would add effort without adding retention. To establish whether these relationships are causal, future work should design interventions that teach users to adopt behaviors linked to better retained performance, either before or during interactions. We differentiate this type of intervention from system-side scaffolding \cite{cohn2026theory,ma2025dbox} because its purpose is to help users develop behavioral patterns that they can carry into future AI use even when such scaffolds are absent. We propose several possible avenues for designing such interventions.

\subsection{Redirecting \actSR{} Toward Active Reasoning} Systems can detect when the user is repeatedly soliciting solutions and then nudge the user toward more explicitly stating their reasoning \cite{hertwig2017nudging}. Repeated, context-sensitive interventions may provide opportunities for users to practice reasoning-preserving interaction strategies, with prior work suggesting that repeated metacognitive questioning can produce effects in learners that persist after intervention removal \cite{bannert2015short}. Future systems could therefore adapt and eventually fade these interventions as users begin to exhibit the targeted behaviors independently. 

Alternatively, a rephrased prompt could be presented to the user, similar to automated prompt enrichment \cite{azad2019query} and recent LLM-based prompt enhancement approaches \cite{sarkar2025conversational,galimzhanova2023rewriting}. Such systems could expose the reformulation to the user so that the intervention also teaches reusable interaction strategies. This type of intervention can occur in the input field or through another interactive companion.

It is important to note, however, that the presence of \actTA{} alone may not indicate sufficient cognitive involvement, as our quantitative subtype analysis reveals. Thus, interventions should not simply teach users to verbalize more reasoning. Rather, future systems could attend to \textit{what role} that articulation plays and teach users interaction strategies that more effectively expose their current reasoning, uncertainty, or misconceptions.

Given that our fine-grained analysis shows that \actTA{} encompasses distinct conversational behaviors, these nudges could attend to \textit{what role} articulation plays in user reasoning, rather than simply increasing its frequency. In particular, verification-style \actTA{} can still elicit direct answers from the assistant, illustrating that two turns labeled as reasoning can allocate very different amounts of cognitive work to the user and the AI. Therefore, simply increasing the frequency of all \actTA{} utterances may not be sufficient, and greater care should be taken to encourage forms of articulation that expose users' underlying reasoning, uncertainty, or misconceptions.

\subsection{Teaching Interaction Strategies That Persist Beyond Scaffolding} Users can also receive interventions that help them adopt behaviors consistent with higher cognitive engagement independently. This parallels AI-literacy interventions: model-specific tutorials can improve reliance calibration \cite{chiang2022exploring}, and scaffolded auditing can prompt more critical scrutiny of outputs \cite{prabhudesai2025here}. In our setting, boosting could explicitly teach users to propose candidate solutions, explain their reasoning, and seek principles or feedback instead of immediately delegating the next step.

Interactive boosting can let users practice these behaviors, while lighter-weight approaches can scale via contrasting examples \cite{booth2013using} or short videos. \citeauthor{cao2025empowering} showed that a lightweight AI-literacy intervention, consisting of five 3-minute videos explaining AI and demonstrating effective vs. problematic ChatGPT use, is capable of improving self-efficacy and knowledge of AI versus a time-matched control \cite{cao2025empowering}. This suggests short, example-based instruction can scale practical interaction strategies.

Building on this, short trainings could contrast ways users participate in problem solving -- for instance, repeatedly asking ``What should I do next?'' versus proposing a solution, justifying it, and asking for critique -- highlighting reusable strategies for retaining more of the reasoning process.

\subsection{Making Reasoning Delegation Visible Through Peer Comparison}
Building on our findings, future systems could help users reflect on how they distribute cognitive work with AI by comparing their conversational behavior with that of peers who demonstrate stronger subsequent unassisted performance. For example, an interface might visualize the proportion of a user's turns containing \actTA{}, \actSR{}, and other dialogue acts alongside the corresponding distributions among relevant peers. Such comparisons could make otherwise difficult-to-observe interaction patterns more salient, helping users recognize when they rely more heavily on direct solution seeking or contribute less of their own reasoning than users who retain stronger performance without assistance. Rather than prescribing a fixed ``optimal'' conversational composition, this feedback provide users a reference point for reflecting upon how much reasoning they retain versus delegate during AI interaction.

\section{Limitations and Future Work}
Our work has several limitations. First, we only measure immediate \textit{near-transfer} to related puzzles of the same game type rather than delayed retention or far transfer, and therefore do not claim durable learning outcomes. Second, Connect Four and Othello provide controlled tasks with objectively optimal solutions, but may not generalize to open-ended domains such as writing or programming, where optimality of assistant answers is not guaranteed. We also deliberately supplied the assistant with the optimal next move to reduce variation in response correctness; this improves experimental control but differs from realistic interactions with AI models. The participants in our studies have been filtered to those capable of producing high-quality prompts; therefore, any generalization to all AI users may be constrained. Finally, our dialogue acts capture \textit{observable} conversational behavior, and the relationships between participant behavior and transfer remain observational. We had no validated pre-test of game skill or of general engagement. The available pre-assistance measures (one unaided move, its confidence rating, and self-rated skill) did not account for the \actTA{} association, but they are coarse, range-restricted by our screening, and weakly predictive of unassisted performance. A pre-test puzzle and a randomized manipulation of \actTA{} are the natural next steps.

Future work should test these patterns across longer timescales, less constrained tasks, and different assistant policies. Additionally, interventions for participants should be designed according to these patterns and examined for effectiveness on improving unassisted performance.

\section{Conclusion}

AI assistance is usually judged by what users accomplish with AI. We ask a different question: what can they do once the AI is gone? Across two controlled problem-solving studies, we find that conversational behavior is meaningfully associated with subsequent independent performance, and that users' conversational choices shape the kind of assistance they receive. Users who maintained cognitive participation during AI assistance performed better independently afterward, though our design cannot establish that this participation produced the better performance. As AI becomes embedded in everyday work, users should learn not only how to succeed with AI support, but also how to retain their ability to reason and perform independently without it.


\bibliographystyle{ACM-Reference-Format}
\bibliography{sample-base}

\appendix

\section{Informed Consent}
\label{app:informed_consent}


This work involves multiple studies with minimally different informed consents. For brevity, we include a parameterized version of the informed consent, and detail how we alter the wording for each study we conducted. 

\begin{tcolorbox}[
  breakable, enhanced,
  colback=gray!3, colframe=gray!55, boxrule=0.5pt, arc=2pt,
  left=8pt, right=8pt, top=8pt, bottom=8pt,
  title={\textbf{Informed Consent}}, fonttitle=\bfseries,
  coltitle=black, colbacktitle=gray!18, label={box:consent}
]
\small
\setlength{\parskip}{4pt}
\setlength{\parindent}{0pt}

\textbf{About this study.}
This is a research study on \texttt{<PURPOSE\_OF\_STUDY>}. Your goal is to play \texttt{<GAME\_COUNT>} \texttt{<GAME>} puzzles.

\textbf{What you will do ($\sim$15 minutes).}
\begin{enumerate}[leftmargin=1.4em,itemsep=1pt,topsep=2pt]
  \item A \textbf{prompt-writing task ($\sim$2 minutes)}.
  \item IF YOUR CREATED PROMPT IS HIGH-QUALITY: Play the first \texttt{<GAME\_1\_DESC>} \texttt{<GAME\_1\_ASSISTANCE>}.
  \item \textbf{Reflection questions} ($\sim$1 minute).
  \item Play \texttt{<GAME\_2\_DESC>} \texttt{<GAME\_2\_ASSISTANCE>}.
  \item \textbf{Reflection questions} ($\sim$1 minute).
\end{enumerate}

\textbf{Voluntary participation.}
Participation is voluntary and you may withdraw any time.

\textbf{Eligibility.}
You must be 18 years of age or older to participate. You must not be a skilled
\texttt{<GAME>} player.

\textbf{Risks \& discomforts.}
Participation in this study involves minimal risk.

\textbf{Benefits.}
There are no direct benefits to you from participating beyond the listed
compensation.

\textbf{Confidentiality \& data retention.}
We collect your Prolific ID, the prompt you create, demographics, \texttt{<AI\_INTERACTION>} and the moves
you make during the game. Data is stored securely and used for research only;
de-identified results may be published.

\textbf{Questions or concerns.}
This study is conducted by \textsc{[Author]}, at \textsc{[Institution]}. If you have
questions about the study, contact \textsc{[email]}.

\medskip \hrule \medskip

$\square$~~I have read and understood the information above, and I consent to
participate in this study.

\medskip
\fbox{\textbf{I agree, continue}}
\end{tcolorbox}

We now discuss the possible configurations for each study in Tables \ref{tab:config_consent_1} and \ref{tab:config_consent_2}. The configurations are composable; a baseline study for Othello applies parameters from both the Baseline row in Table \ref{tab:config_consent_1} and the Othello row in Table \ref{tab:config_consent_2}.

\begin{table}[t]
  \centering
  \small
  \begin{tabularx}{\columnwidth}{@{}XXXXX@{}}
    \toprule
    \textbf{Configuration} & \texttt{<PURPOSE\_OF\_STUDY>} & \texttt{<GAME\_1\_ASSISTANCE>} & \texttt{<GAME\_1\_ASSISTANCE>} & \texttt{<AI\_INTERACTION>} \\
    \midrule
    Baseline &  How \textbf{users with different AI usage skills} make decisions while playing games & N/A & N/A & N/A\\
    \midrule
    Assisted & How \textbf{proficient AI users} make decisions while playing AI-assisted games, so that we can help less-experienced users on similar tasks. & , with AI assistance & , without AI assistance & your conversation with the AI, \\
    \bottomrule
  \end{tabularx}
  \caption{Configurations of AI-usage parameters in the informed consents for our different studies.}
  \label{tab:config_consent_1}
\end{table}

\begin{table}[t]
  \centering
  \small
  \begin{tabularx}{\columnwidth}{@{}lllXX@{}}
    \toprule
    \textbf{Configuration} & \texttt{<GAME>} & \texttt{<GAME\_COUNT>} & \texttt{<GAME\_1\_DESC>} & \texttt{<GAME\_2\_DESC>} \\
    \midrule
    Connect Four & Connect Four & two & Connect Four puzzle (6 minutes) & the second game (3 minutes) \\
    \midrule
    Othello & Othello & three & Othello game (8 minutes) & a round of two games (3 minutes) \\ 
    \bottomrule
  \end{tabularx}
  \caption{Configurations of game parameters in the informed consents for our different studies.}
  \label{tab:config_consent_2}
\end{table}

\section{Puzzle Configurations}

Table~\ref{tab:puzzle_time_settings} summarizes the puzzles and allotted completion times.

\begin{table}[!h]
    \centering
    \begin{tabular}{l|lll}
    \toprule
        \textbf{Game} & \textbf{Assisted (Time)} & \textbf{Unassisted 1 (Time)} & \textbf{Unassisted 2 (Time)}  \\
        \midrule
        Connect Four & \href{https://sites.math.rutgers.edu/~zeilberg/A4/ch5/P15.html}{\underline{Win-in-5-moves}} (6 mins)& \href{https://sites.math.rutgers.edu/~zeilberg/A4/ch4/P6.html}{\underline{Win-in-4-moves}}(3 mins)& N/A \\
        \midrule
        Othello & \href{https://othelloclub.com/en/board.php?board=0222222021221201011222211121222111221221201221210022222100021221&turn=1&date=20260727}{\underline{Expert Puzzle}} (8 mins) & \href{https://othelloclub.com/en/board.php?board=2211120021111100211211112221211122211111211121112111210111111111&turn=2&date=20260706}{\underline{Beginner Puzzle \#1}} (1.5 mins) & \href{https://othelloclub.com/en/board.php?board=2111111120112222211121212112121121212111211211112111210011111100&turn=2&date=20260726}{\underline{Beginner Puzzle \#2}} (1.5 mins) \\
        \bottomrule
    \end{tabular}
    \caption{The different puzzle tasks and time allotted for our game studies. We aimed to match the puzzles by selecting unassisted puzzles with motifs similar to those of the assisted puzzle, without sharing the initial optimal moves.}
    \label{tab:puzzle_time_settings}
\end{table}

\section{Prompt Creation Screener Details}
\label{app:prompt_creation}

To filter participants based on their prompt construction capability, which is our proxy for Generative AI interaction literacy, we filter participants through a prompt creation task. Specifically, we present participants with the following scenario:

\begin{quote}
    \begin{itemize}
        \item \textit{Julie posts cooking videos to 1.2M followers who love her quick, healthy recipes.}
        \item \textit{Over the next three years she wants to grow online and publish a cookbook (torn on traditional vs. self-publishing).}
        \item \textit{She works four days weekly, limited time and budget.}
    \end{itemize}
\end{quote}

We instruct participants to help Julie devise a concrete three-year plan. They can construct a single prompt to a state-of-the-art AI agent that must produce a deliverable. 

After constructing this prompt, the final prompt is sent to a panel of three LLM judges for a more reliable assessment \cite{li2024llms}. The prompt quality is scored on a scale from 1 (poor) to 4 (excellent). The three LLM judges are \texttt{GPT-5.4-mini} with chain-of-thought reasoning, \texttt{Llama-3.3-70B-Instruct} \cite{grattafiori2024llama}, and \texttt{Claude-sonnet-5}. When selecting models, we aimed for low latency and decent agreement with humans. The final score of the prompt is aggregated by averaging the individual scores. See the rubric below:

\begin{quote}
    Score 1 (Poor / off-target): Vague or minimal; reads like a quick search query; omits most of the scenario.
    
    Score 2 (Developing): Conveys the gist with several goals/constraints but loosely; incomplete coverage and little structure.
    
    Score 3 (Proficient): Captures essentially all key scenario elements clearly; a complete, competent brief, even if plainly formatted.
    
    Score 4 (Excellent): Proficient PLUS deliberate prompt craft — explicit structure (sections/roles/output spec) and/or meta-instructions (self-check, clarifying questions, reasoning guidance) beyond merely restating the scenario.
\end{quote}

To assess the alignment between LLM and human judgment, we collected a set of 34 prompts from a previous round of the prompt creation task with a longer fictional scenario but the same rubrics. One of the authors manually scored each of the prompts along the rubric. The panel's mean score correlates strongly with the author's ($r = 0.829$, $p < .001$; Spearman $\rho = 0.829$; ICC(2,1) $= 0.787$), and the panel's rounded score---the one the screener acts on---agrees with the author's at a quadratic-weighted Cohen's $\kappa$ of $0.712$: the two assign the same score to $59\%$ of prompts and land within one point on $97\%$. We acknowledge that the agreement is not perfect, but is sufficient for screening participants, as the panel on average tends to score prompts lower than humans (2.47 vs.\ 2.53 on the prompt set; Wilcoxon signed-rank $p = .69$, so the two do not differ systematically). Because these 34 prompts come from an earlier version of the prompt task with a longer fictional scenario, this agreement estimate should be read as indicative for the deployed task rather than as a direct measurement of it.

\section{Connect Four Study: Additional Results}

\subsection{Demographic Differences}
\begin{figure}
    \centering
    \includegraphics[width=\linewidth]{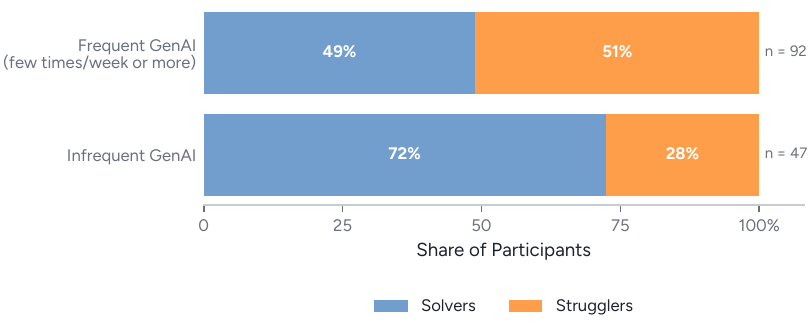}
    \caption{Proportions of Solvers vs. Strugglers among frequent and infrequent Generative AI users.}
    \Description{Two horizontal stacked bars, each spanning the full width as 100
    percent of a group. The upper bar, ``Frequent GenAI (few times per week or more)'',
    $n = 92$, is split 49 percent Solvers in blue and 51 percent Strugglers in orange.
    The lower bar, ``Infrequent GenAI'', $n = 47$, is split 72 percent Solvers and 28
    percent Strugglers. Infrequent users are therefore the more successful group.}
    \label{fig:demo_diff}
\end{figure}

From preliminary observations, we identify \textbf{frequent users of generative AI} - those who use Generative AI more than a few times a week - are disproportionately Strugglers (Figure \ref{fig:demo_diff}). We test this observation and find that the differences are significant ($\chi^2 = 9.49,\ p < .01$). While this pattern may be consistent with over-reliance on AI, our study design does not establish why prior GenAI usage may affect transfer performance. We therefore treat this result as exploratory.

\subsection{Conversation Statistics}
On average, Solvers conversed with the AI assistant in the assisted puzzle for 2.5 turns, compared to 2.2 turns for Strugglers. Solvers used more unique words compared to Strugglers (22 vs. 17) on average. These differences are not statistically significant. 

\subsection{Self-report Survey Results}

\paragraph{AI Assistant Survey} Both Solvers and Strugglers find the assistant capable of providing responses consistent with their intent and easy to use. These ratings do not differ significantly between Solvers and Strugglers (Table \ref{tab:ai_assistant_rating}).

\begin{table}[]
    \centering
    \begin{tabular}{lcc}
    \toprule
        \textbf{Item} & \textbf{Solvers} & \textbf{Strugglers} \\
        \midrule
        Consistent with Intent & 4.32 $\pm$ 0.17 & 4.08 $\pm$ 0.27 \\
        Easy to Use & 4.23 $\pm$ 0.20 & 4.32 $\pm$ 0.19 \\
        \bottomrule
    \end{tabular}
    \caption{Average Likert scale ratings $\pm$ 95\% CI from 1 (Strongly Disagree) to 5 (Strongly Agree), scored by Solvers and Strugglers along two AI assistant evaluation axes.}
    \label{tab:ai_assistant_rating}
\end{table}

\begin{figure}
    \centering
    \includegraphics[width=\linewidth]{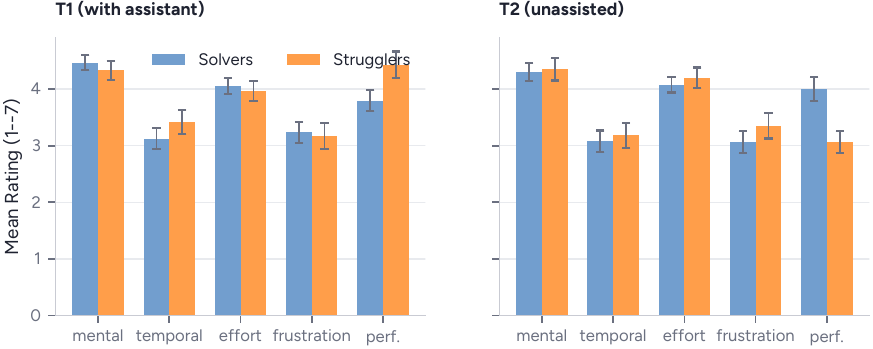}
    \caption{Average NASA-TLX ratings with error bars indicating 95\% CI, across two puzzles in the Connect Four trials.}
    \Description{Two panels of grouped bar charts, one per Connect Four round, with
    five NASA-TLX items along the horizontal axis---mental demand, temporal demand,
    effort, frustration and performance---and mean rating on a 1 to 7 scale on the
    vertical axis, with error bars showing 95 percent confidence intervals. Blue bars are
    Solvers ($n = 79$) and orange bars Strugglers ($n = 60$). In the assisted round the
    two groups are close on every item, the largest gap being the performance item, where
    Strugglers rate themselves 4.43 against the Solvers' 3.80. In the unassisted round
    that gap reverses and widens: Solvers rate their performance 4.00 against the
    Strugglers' 3.07, while the other four items remain within about 0.3 of one another.
    The performance item is worded so that a higher rating means the participant felt
    more successful.}
    \label{fig:cf_nasa_tlx}
\end{figure}

\paragraph{NASA-TLX} Participants tend to find both Connect Four tasks challenging, as they experience high mental load and exert significant effort across both rounds. Interestingly, Strugglers self-perceive to attain higher performance in the AI-assisted round, and this trend reverses for the unassisted round (Figure \ref{fig:cf_nasa_tlx}). This is consistent with the pattern that Strugglers can obtain better P1 performance by delegating to the AI assistant completely, but would fail to achieve similar performance on the unassisted puzzle.

\definecolor{promptlive}{HTML}{2B6349}
\definecolor{promptcode}{HTML}{729ECE}
\newtcolorbox{promptbox}[2][promptcode]{%
  enhanced, breakable, colback=#1!4!white, colframe=#1!55!white, boxrule=0.6pt,
  arc=2pt, left=7pt, right=7pt, top=6pt, bottom=6pt,
  fonttitle=\bfseries\small, coltitle=white, colbacktitle=#1,
  title={#2}, fontupper=\footnotesize\ttfamily}

\section{Study Interfaces}
\label{app:interfaces}

\begin{figure*}[t]
  \centering
  \begin{tabularx}{\textwidth}{@{}XX@{}}
    \textbf{\small Connect Four} & \textbf{\small Othello} \\[2pt]
    \includegraphics[width=\linewidth]{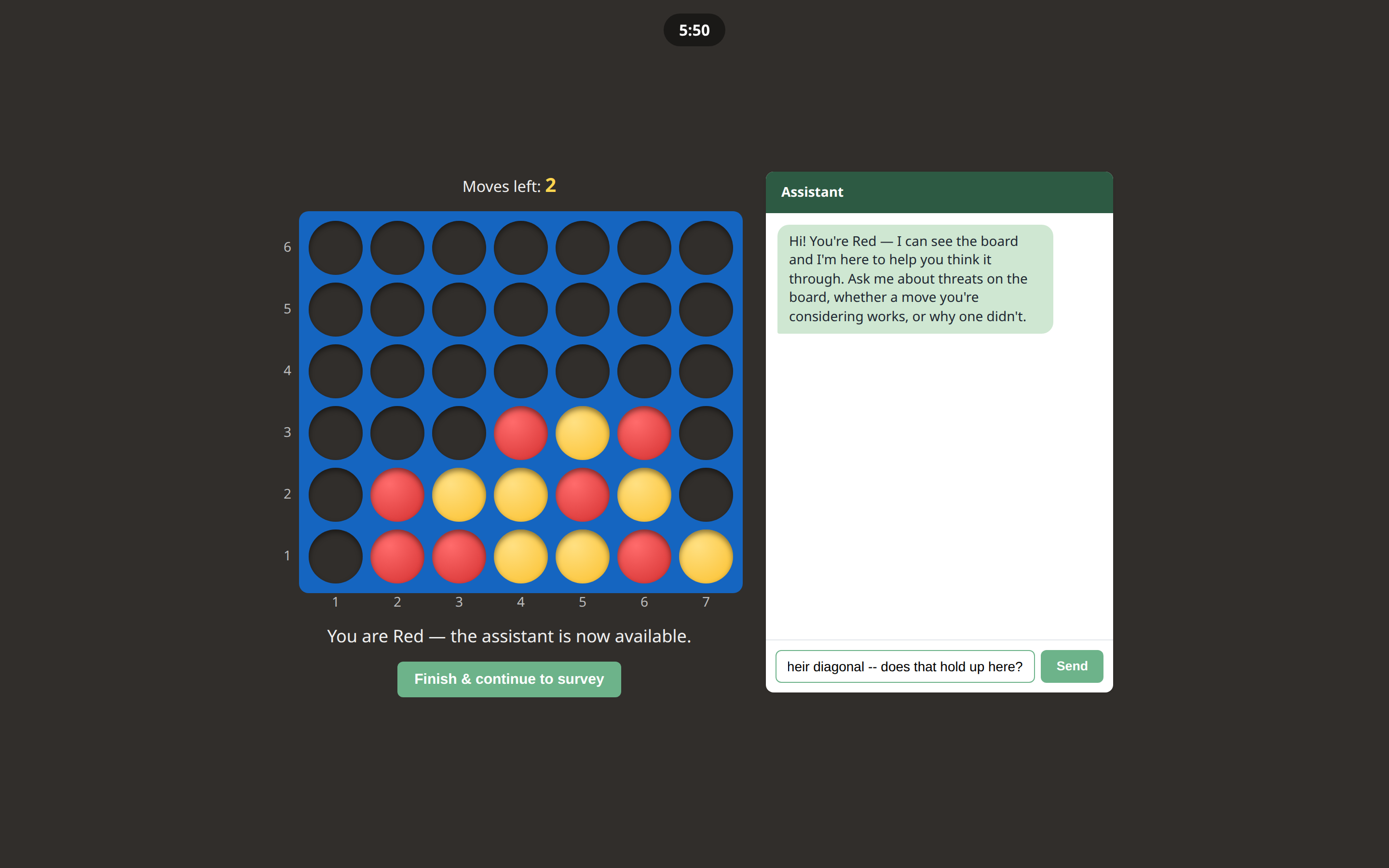} &
    \includegraphics[width=\linewidth]{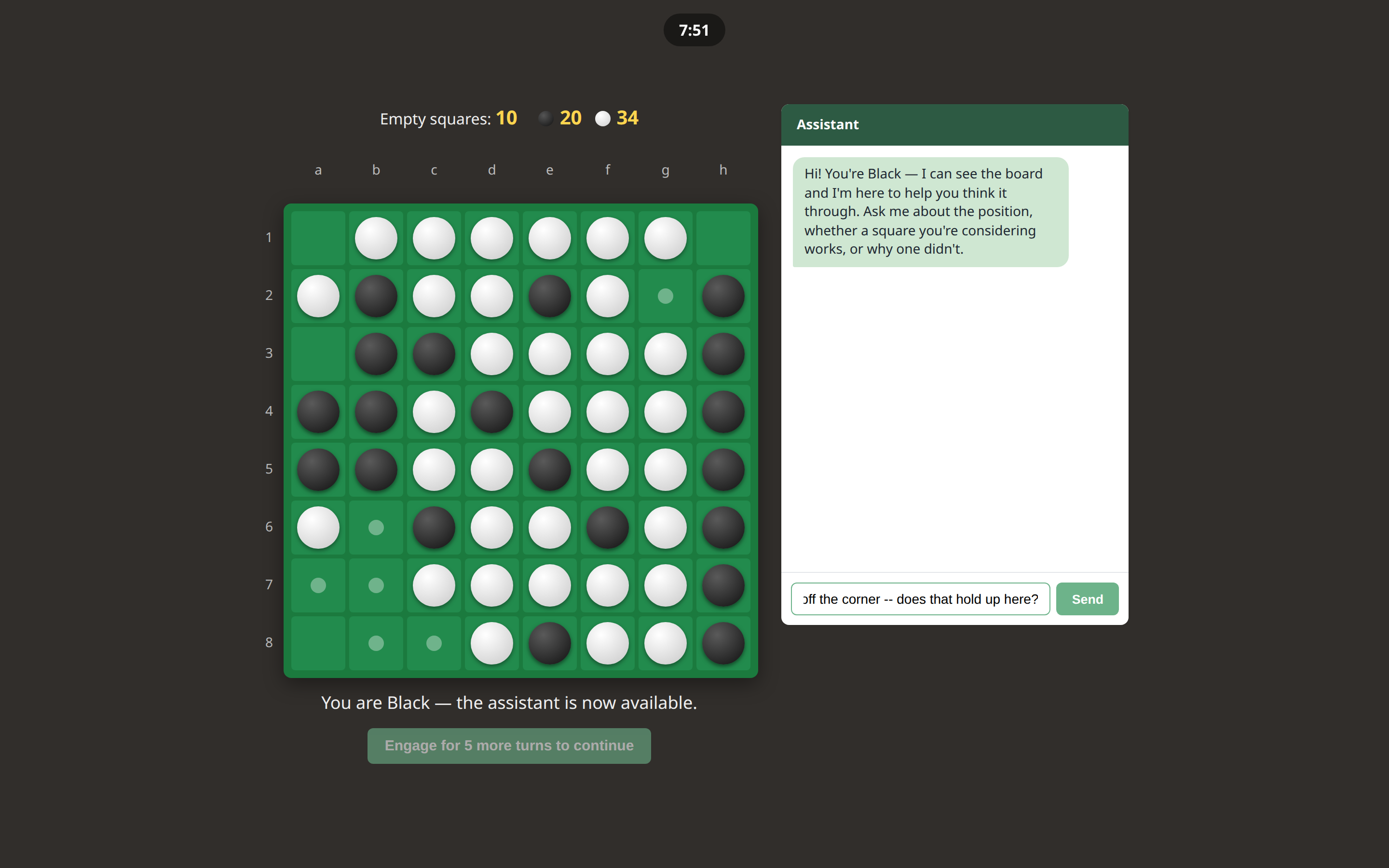} \\[6pt]
    \multicolumn{2}{@{}l@{}}{\textbf{\small The prompt-writing screener and the in-game confidence gate}} \\[2pt]
    \includegraphics[width=\linewidth]{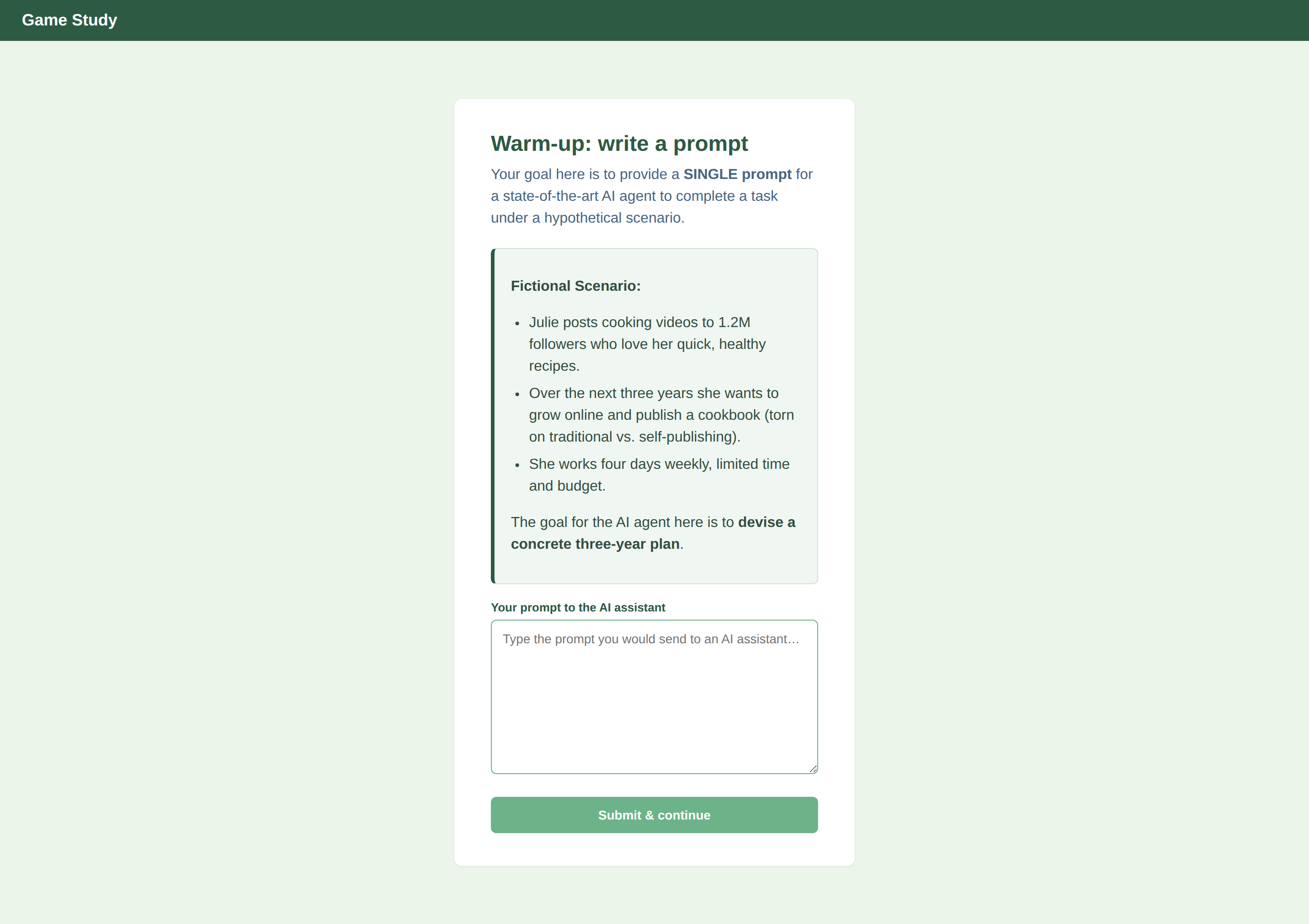} &
    \includegraphics[width=\linewidth]{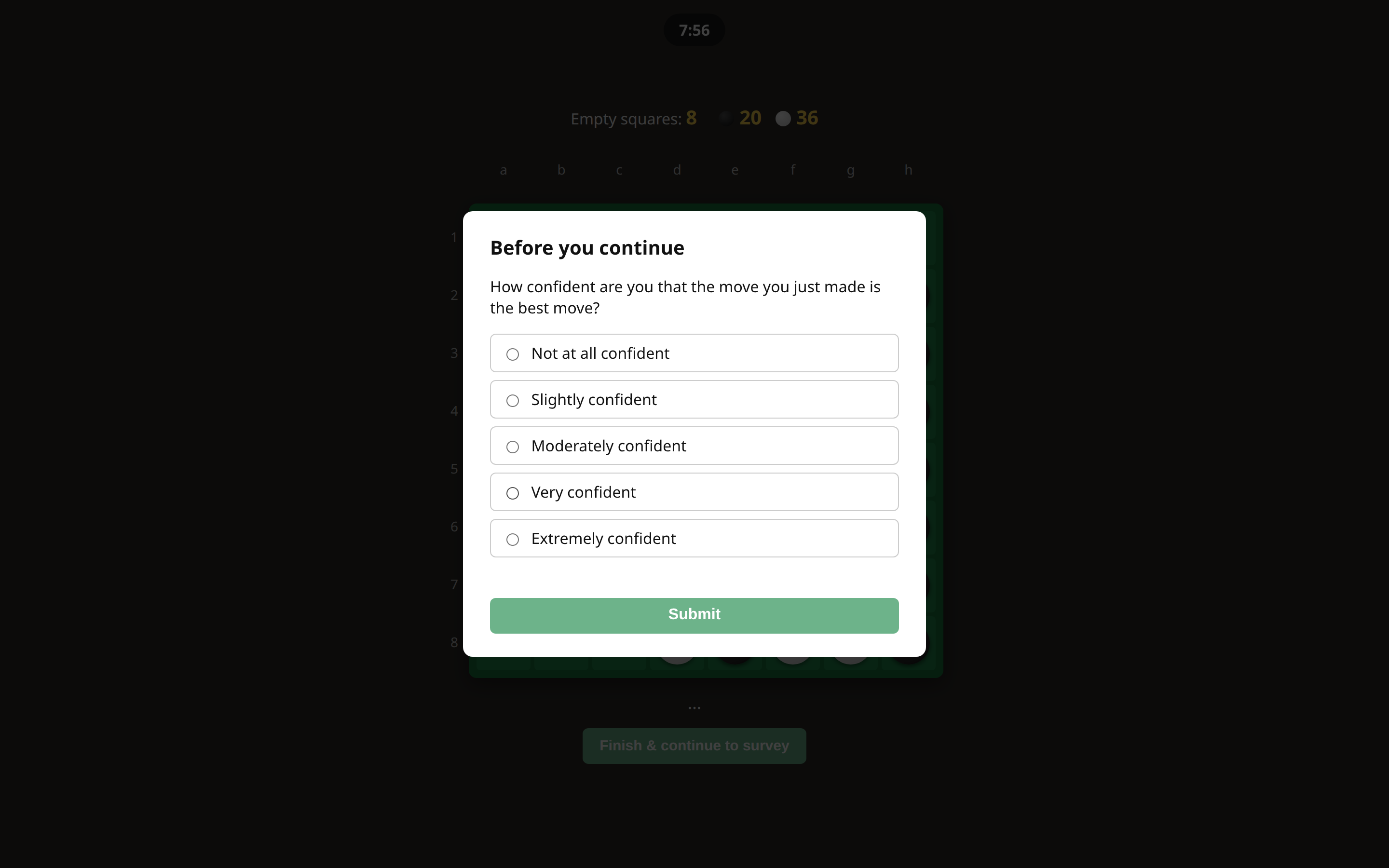} \\
  \end{tabularx}
  \caption{The study interface. Top: the assisted round in each game, with the board
  beside the assistant panel and the engagement requirement shown beneath. Bottom left:
  the prompt-writing screener, scored by the LLM judge panel described in
  Appendix~\ref{app:prompts}. Bottom right: the confidence gate, which fires after the
  participant's first unaided move.}
  \Description{Four screenshots of the study web interface. The top two show the
  assisted round: on the left the Connect Four board beside a chat panel headed ``AI
  assistant'' with an opening message and a text box, and a disabled button reading
  ``Engage for 5 more turns to continue''; on the right the same layout with an Othello
  board. The bottom left screenshot shows the prompt-writing task, a scenario
  description above a large empty text area. The bottom right shows the confidence
  gate: a dialogue over the dimmed board asking ``How confident are you that the move
  you just made is the best move?'' with five radio options from ``Not at all
  confident'' to ``Extremely confident'' and a Submit button.}
  \label{fig:interfaces}
\end{figure*}

Figure~\ref{fig:interfaces} shows the screens participants saw. Both studies used the
same instrument, differing only in the game: an eligibility and consent flow, a
prompt-writing screener, a rules primer, the timed puzzle, and a post-round survey.
The assistant panel is hidden until the participant has made one unaided move and
answered the confidence gate, so the first move in the assisted round is always their
own.

\section{Othello Study: Additional Results}
\label{app:othello_extra}

Participants are split at the median of unassisted performance (28.5 discs; $n = 61$
at or above, $n = 50$ below). These comparisons were not prespecified and are reported
as exploratory: p-values are uncorrected across the fourteen tests below.

\subsection{Conversation Statistics}

The two groups talk to the assistant for the same number of turns --- 5.54 against 5.56
($U = 1642$, $p = .411$) --- which is unsurprising, since five turns were required. They
differ in what those turns contain: participants at or above the median use more
distinct words (41.5 against 33.6; $U = 1944$, $p = .013$) and write longer turns (10.2
against 8.1 words). The engagement floor therefore equalises conversation
\emph{length} while leaving conversation \emph{content} free to vary, which is the
variation the paper's analysis rests on.

\subsection{Self-report Survey Results}

\paragraph{AI Assistant Survey} Participants rated the assistant favourably overall
(Table~\ref{tab:oth_ai_rating}). Those who went on to transfer better rated it
\emph{less} easy to use ($4.03$ against $4.39$, $U = 1129$, $p = .013$), while ratings
of intent-consistency did not differ ($p = .622$).

\begin{table}[t]
    \centering
    \begin{tabular}{lccc}
    \toprule
        \textbf{Item} & \textbf{At/above median} & \textbf{Below median} & \textbf{All} \\
        \midrule
        Consistent with Intent & 4.03 $\pm$ 0.27 & 4.20 $\pm$ 0.24 & 4.10 \\
        Easy to Use            & 4.03 $\pm$ 0.22 & 4.39 $\pm$ 0.23 & 4.19 \\
        \bottomrule
    \end{tabular}
    \caption{Average Likert ratings $\pm$ 95\% CI, from 1 (Strongly Disagree) to 5
    (Strongly Agree), by unassisted-performance group in the Othello study
    ($n = 109$ of 111 responded).}
    \label{tab:oth_ai_rating}
\end{table}

\paragraph{NASA-TLX} During the \emph{assisted} round, participants who went on to
transfer better reported higher effort ($4.90$ against $4.34$, $p = .045$) and rated
their own performance markedly \emph{lower} ($2.72$ against $3.66$, $p = .012$). In the
final unassisted round that performance gap reverses ($3.23$ against $2.64$,
$p = .054$), and frustration trends the other way ($3.95$ against $4.42$, $p = .057$);
mental and temporal demand do not differ in either round. The pattern mirrors the
Connect Four result in Figure~\ref{fig:cf_nasa_tlx}: the participants who felt least
successful while assisted are the ones who did best once the assistant was removed.

\subsection{Robustness Check for Othello's Participant-side Dialogue Act Results}
\label{app:othello_robustness}
As a robustness check, we model transfer performance---the participant's own final disc count averaged over $O_{U1}$ and $O_{U2}$---using ordinary least squares (OLS) regression. We include prior AI usage frequency, assisted-task performance ($O_{A}$ disc count), and \actSR{} DAP as covariates (Table~\ref{tab:ols_regression_othello}). The TA-only specification has the lowest BIC, and the \actTA{} association remains stable after adding AI usage, \actSR{} prevalence, and assisted-round performance. In contrast, \actSR{} DAP shows no significant relationship with transfer once \actTA{} is included. \actTA{} and \actSR{} are themselves negatively correlated ($r=-0.26$), suggesting that the two measures capture partially overlapping interaction patterns. Across both games, participants who more frequently articulate their own reasoning while working with the assistant perform better on subsequent unassisted tasks. 

\begin{table}[]
    \centering
    \begin{tabular}{c|cccc|c}
    \toprule
       \textbf{Model} & $\beta_{\text{TA-DAP}}$ & $\beta_{\text{AI}}$ & $\beta_{\text{SR-DAP}}$ & $\beta_{O_{A}}$  & \textbf{BIC} \\
       \midrule
        \textbf{+ TA-DAP }& \textbf{+0.375**} & -- &  -- & -- & \textbf{303.9} \\
        + AI & +0.380** & -0.131 & -- & -- & 306.4 \\
        + SR-DAP & +0.334** & -0.132 & -0.180 & -- & 307.1 \\
        + $O_{A}$ & +0.334** & -0.136 & -0.154 & -0.081 & 311.0 \\
        \bottomrule
    \end{tabular}
    \caption{OLS regression results for the Othello experiments, modeling the mean unassisted disc count as the dependent variable, with different combinations of covariates. Both the dependent variable and the predictors are standardized, so coefficients are in standard-deviation units. All models are fit on the same $n = 110$; one participant whose assisted round timed out before any move is excluded. **$p < .01$}
    \label{tab:ols_regression_othello}
\end{table}

\paragraph{Pre-assistance skill.} $O_{A}$ is scored with the assistant available, so the model above does not separate prior ability from the conversation's contribution. We therefore added three measures taken before the assistant unlocked, recorded in the in-game gate (Section~\ref{sec:study_procedure}, step 3) and at eligibility: whether the unaided first move was optimal (23\% of participants), confidence in that move (1--5), and self-rated Othello skill (1--5, with ``I do not know this game'' coded 0). None correlates with \actTA{} DAP (Spearman $\rho = +0.02$ to $+0.14$, all $p \ge .13$) or with transfer ($\rho = +0.05$ to $+0.15$, all $p \ge .12$). The \actTA{} coefficient is stable when they are added (Table~\ref{tab:prior_skill_othello}). The partial Spearman correlation is likewise stable: $\rho = +0.358$, $+0.353$ and $+0.344$ holding each measure in turn, against $+0.358$ unadjusted.

These measures are weak proxies for the skill that matters here. In the Baseline condition ($n = 100$), where no one had an assistant, a correct first move predicts the gated puzzle's own outcome ($\rho = +0.50$). That is partly mechanical, since a correct first move is kept on the board. It does not predict the two later puzzles ($\rho = -0.14$, $p = .16$), and neither confidence nor self-rated skill predicts either outcome (all $|\rho| \le 0.14$). Range restriction contributes. Self-rated skilled players were screened out at eligibility, and correct first movers reporting high confidence were screened out in-game, so no retained correct first mover rated confidence above 3. The result therefore excludes confounding by entry-level skill as these items capture it. It cannot exclude a more general disposition to attend to the board, which would require a validated pre-test or a manipulation of \actTA{} itself.

\begin{table}[]
    \centering
    \small
    \begin{tabular}{l|cc|c}
    \toprule
       \textbf{Model} & $n$ & $\beta_{\text{TA-DAP}}$ [95\% CI] & \textbf{BIC} \\
       \midrule
        TA-DAP only & 110 & +0.375*** [+0.20, +0.55] & 303.9 \\
        + pre-assistance skill & 110 & +0.352*** [+0.17, +0.53] & 314.8 \\
        \midrule
        Full model (Table~\ref{tab:ols_regression_othello}) & 110 & +0.334*** [+0.15, +0.51] & 311.0 \\
        + pre-assistance skill & 110 & +0.303** [+0.12, +0.49] & 321.4 \\
        + NASA-TLX effort$^{\dagger}$ & 110 & +0.299** [+0.11, +0.49] & 326.0 \\
        \bottomrule
    \end{tabular}
    \caption{\actTA{} DAP and Othello transfer, adjusting for skill measured before the assistant was available: first-move correctness, first-move confidence, and self-rated skill, all standardized. No pre-assistance term is significant (all $p \ge .13$). The full model adds AI usage, \actSR{} DAP and $O_{A}$, as in Table~\ref{tab:ols_regression_othello}. $^{\dagger}$Effort is reported after the assisted round and correlates with \actTA{} DAP ($\rho = +0.19$, $p = .04$), so it is post-treatment and included only as a sensitivity check. **$p < .01$, ***$p < .001$}
    \label{tab:prior_skill_othello}
\end{table}

\section{Is \texttt{Think Aloud} Prevalence Just Verbosity?}
\label{app:verbosity}

\actTA{} prevalence could serve as a proxy for talking more: a participant who writes longer messages has more room to articulate reasoning and would accumulate \actTA{} labels without the act itself conveying anything. We therefore compared it against three measures of sheer volume: the number of turns a participant took, the total words they wrote, and their mean words per turn.

Volume does correlate with transfer, but weakly, and \actTA{} prevalence is not reducible
to it. Against the averaged unassisted disc count ($n = 111$), \actTA{} DAP reaches
$\rho = +0.358$, while mean words per turn reaches $+0.246$ ($p = .009$), total words
$+0.210$ ($p = .027$), and turn count $+0.075$ ($p = .44$). The three volume measures are
themselves correlated with \actTA{} DAP ($\rho = +0.44$ to $+0.46$ for the two word
measures), so the question is which one carries the relationship. Partialling settles it,
and asymmetrically: holding turn count, total words, or mean words per turn in turn,
\actTA{} DAP stays at $\rho = +0.352$, $+0.302$ and $+0.288$ (all $p \le .002$), whereas each
volume measure falls to nothing when \actTA{} DAP is held ($\rho = +0.005$, $+0.056$ and
$+0.107$; all $p > .26$). Adding mean words per turn to the full model in
Table~\ref{tab:ols_regression_othello} leaves \actTA{} DAP at $+0.329$ ($p < .01$) and the
new term at $+0.011$, and raises BIC, so verbosity earns no place in the model.

The same holds in Connect Four, at the weaker effect size that study supports: \actTA{}
DAP reaches $\rho = +0.182$ ($p = .032$) against the unassisted score while turn count,
total words and mean words per turn are all null ($+0.044$, $+0.080$, $+0.061$), and
\actTA{} DAP is essentially unchanged when each is held ($+0.165$ to $+0.177$).

The clearest evidence is a dissociation between the two outcomes. In Othello, turn count
is related to the \textit{assisted} round ($\rho = +0.261$, $p = .008$) but not to
transfer ($+0.075$), while \actTA{} DAP is the opposite: unrelated to the assisted
round ($-0.052$, $p = .60$) and strongest on transfer. As in
Table~\ref{tab:dissociation}, assisted-round correlations use the 101 participants whose
assisted round ended in a completed game rather than the timer; including the 10
timed-out rounds ($n = 111$) leaves the pattern intact (turn count $+0.198$, $p = .037$;
\actTA{} DAP $-0.093$, $p = .33$). Were \actTA{} DAP volume in disguise, it
would behave like turn count on both outcomes rather than its opposite. One qualification:
the \textit{count} of \actTA{} turns also tracks transfer ($+0.327$, $p < .001$), so what
the data separate is reasoning talk from talk in general, not a rate from an amount.

\section{LLM Simulations}
\subsection{Agentic Simulation of Full Participant Trajectories}
\label{app:agentic_sim}

We aimed to increase the number of analyzable conversations by directly simulating participants using LLMs. To achieve sufficient diversity in the simulated behavior, we use models from multiple providers and sample diverse personas from Persona-Hub \cite{ge2024scaling}, as conditioning on personas has been shown to influence model behavior. Specifically, we first perform k-means clustering with 60 clusters to identify the main archetypes among the 1M personas, and then select the cluster centers as the representatives.

The simulated participants are instantiated as agents with three different tools: \texttt{view\_board}, \texttt{play\_move}, and \texttt{consult\_assistant}. Similar to Othello phase participants, the agent must interact with the AI assistant for 5 turns. All other conditions are held consistent for the simulations, from the specific puzzles to the scoring mechanisms. Furthermore, to make simulated utterances more human-like to elicit similar distributions in assistant utterances, we incorporate 14 in-context human utterance examples spanning different dialogue acts. The in-context examples increase the stylistic similarity between real and simulated utterances from 0.88 to 0.97, as measured by cosine similarity using StyleDistance embeddings \cite{patel2025styledistance}. In total, we produce 270 conversations across different persona-model-prompt combinations and 1347 turns. 

\begin{table}[t]
\centering

\begin{tabular}{l r@{\,}l r@{\,}l r@{\,}l}
\toprule
 & \multicolumn{6}{c}{Participant act} \\
\cmidrule(l){2-7}
Assistant act & \multicolumn{2}{c}{Think Aloud} & \multicolumn{2}{c}{Solution Request} & \multicolumn{2}{c}{Knowledge Deficit Q.} \\
\midrule
Board Report &  0.54 & *** & 0.41 & *** & 4.14 & *** \\
Provide Correct Answer &  2.22 & *** & 6.07 & *** & 0.07 & *** \\
\bottomrule
\end{tabular}

\caption{Odds ratio for each assistant act pooled over all \textit{simulated} turns ($n = 1347$). Values above 1 indicate the assistant act is more likely; below 1, less likely.
$^{***}p<.001$}
\label{tab:sim-assistant-acts-pooled}
\end{table}

We report odds ratios for the assistant dialogue acts of interest, given that specific simulated participant acts are present in the previous turn (Table \ref{tab:sim-assistant-acts-pooled}). Consistent with human participants, Solution Request strongly elicits Provide Correct Answer. Similarly, the temporal orientation correlation with Board Report reproduces ($OR = 6.940, p < .001$). There are, unsurprisingly, still discrepancies between human and simulated behaviors: Think Aloud co-occurs with a higher rate of Provide Correct Answer rather than suppressing it, and the presence of Knowledge-Deficit Questions appears to be tied to Board Report. This likely stems from qualitative differences between human and simulated participants: LLMs are more likely to include specific board coordinates in their queries, and there are fewer single-act turns by simulated participants. The simulation results therefore reproduce the \actSR{} and temporal-orientation patterns we discover in human participants, but not the \actTA{} pattern. Because full simulated trajectories diverge from human conversations in these ways, we test the causal link between participant and assistant acts with utterance-level counterfactual replay of real participant turns instead (Appendix~\ref{app:controlled_replay}).

\subsection{Controlled Replay of Participant Utterances through LLM Simulations}
\label{app:controlled_replay}

The previous studies focusing on human-AI dialogue revealed correlational relationships between participants' and assistants' speech acts, but do not establish causal relationships between them. As clean manipulation of human behavior is challenging \cite{kuhlen2013language}, we leverage LLM simulations. Rather than simulate complete participant trajectories \cite{hu2024quantifying,anthis2025llm,park2023generative}, which produced interactions that differed from human conversations (Appendix~\ref{app:agentic_sim}), we use controlled counterfactual replay at the utterance level.

Our approach involves intervening at turn granularity while \textbf{fixing the conversation history, board state, and the assistant}. We perform targeted modifications of natural participant utterances occurring \textit{during gameplay} to produce minimal pairs using \texttt{gpt-5.6-luna}. We focus on the following rewrites:
\begin{enumerate}
    \item Solution Request $\leftrightarrow$ Move Candidate Evaluation (Common Ground Question).
    \item Retrospective $\leftrightarrow$ Prospective.
\end{enumerate}

Examples of minimal pairs are in Table \ref{tab:minimal-pairs}. We constrain our pool of participant utterances to those made during the game to ensure that all four types of utterances we study are plausible. Both members of the minimal pair are then presented to the assistant, and we observe the resulting differences.

\begin{table*}[t]
\centering
\small
\begin{tabularx}{\textwidth}{@{}l X X@{}}
\toprule
Rewrite & Original participant utterance & Minimally edited counterfactual \\
\midrule
prosp.\ $\rightarrow$ retro.
  & \textit{``What would you recommend now?''} & \textit{``How did b7 turn out?''} \\
  & \textit{``where should i go then''}        & \textit{``where did b7 go then''} \\
\addlinespace
retro.\ $\rightarrow$ prosp.
  & \textit{``was that correct?''}             & \textit{``should i play a7 next?''} \\
\addlinespace
cand.\ eval $\rightarrow$ sol.\ req.
  & \textit{``Would a7 be a good move now?''}  & \textit{``What would be a good move now?''} \\
  & \textit{``i dont see much better. a7?''}   & \textit{``i dont see much better. what do you recommend?''} \\
\addlinespace
sol.\ req.\ $\rightarrow$ cand.\ eval
  & \textit{``Okay what should I do next?''}   & \textit{``Okay should I play a7 next?''} \\
  & \textit{``what is the best move then?''}   & \textit{``is a7 the best move then?''} \\
\bottomrule
\end{tabularx}
\caption{Example minimal pairs used for establishing causal claims.}
\label{tab:minimal-pairs}
\end{table*}

\textbf{Rewriting a single participant utterance shifts the assistant's next act in the predicted direction} (Table \ref{tab:replay-targets}). Re-orienting a prospective question toward the move just played
raises Board Report from 16\% to 94\% ($+78$ pp, $p < .001$). Transforming a Move Candidate Evaluation into a Solution Request raises Provide Correct Answer from 46\% to 94\% ($+48$
pp, $p < .001$); rewriting in the opposite direction lowers it from 94\% to 74\% ($-20$ pp,
$p < .001$). Because the outcome is a paired binary indicator --- whether the assistant's
next turn carries the target act, before and after the rewrite --- significance is by exact
McNemar test on the discordant pairs. Over the full set of human-sourced stimuli the
manipulations are close to one-sided: re-orienting toward the move just played adds Board
Report in $86$ pairs and removes it in $1$ ($n = 131$, $p < .001$); rewriting a candidate
evaluation as a solution request adds Provide Correct Answer in $155$ pairs and removes it
in $10$ ($n = 343$, $p < .001$); the reverse rewrite removes it in $98$ pairs and adds it in
$14$ ($n = 384$, $p < .001$). 

\begin{table}[t]
\centering
\small
\begin{tabular}{@{}l r r r r@{\,}l@{}}
\toprule
Rewrite & $n$ & orig. & rewr. & \multicolumn{2}{c}{$\Delta$ (pp)} \\
\midrule
\multicolumn{6}{@{}l}{\emph{Orientation}\quad target act: Board Report} \\
\quad prosp.\ $\rightarrow$ retro. &  32 & 16\% & 94\% & $+78$ & *** \\
\quad retro.\ $\rightarrow$ prosp. & \multicolumn{5}{l}{$n=3$ --- too few to test} \\
\addlinespace
\multicolumn{6}{@{}l}{\emph{Request type}\quad target act: Provide Correct Answer} \\
\quad cand.\ eval $\rightarrow$ sol.\ req. & 149 & 46\% & 94\% & $+48$ & *** \\
\quad sol.\ req.\ $\rightarrow$ cand.\ eval & 132 & 94\% & 74\% & $-20$ & *** \\
\bottomrule
\end{tabular}
\caption{Participant utterance manipulation results. Columns give the percentage of
assistant responses containing the target act before and after rewriting, and the paired difference in percentage points (pp). The outcome is paired and binary, so significance is by exact McNemar test on the discordant pairs.
$^{*}p<.05$, $^{***}p<.001$.}
\label{tab:replay-targets}
\end{table}

We find that \textbf{these causal relationships replicate} when we replay the same stimuli (system prompt, conversation, and board state) \textbf{across four assistant models} from three providers (Table~\ref{tab:coach-models}). The magnitude of changes given a dialogue act, however, is constrained by the model's general tendency to respond with that dialogue act (Table \ref{tab:coach-baselines}).

Table~\ref{tab:coach-baselines} gives each replay assistant's rate of producing the two target
acts on the \emph{unmodified} participant utterance, pooled over all human stimulus pairs. These
rates bound the headroom available to a rewrite and account for the spread in magnitude reported
in Table~\ref{tab:coach-models}: an assistant that already volunteers the answer in most replies
has little room left to move, while one that rarely does has little propensity to amplify.

\begin{table}[t]
\centering
\small
\begin{tabular}{@{}l r r r r@{\,}l@{}}
\toprule
Coach model & $n$ & orig. & rewr. & \multicolumn{2}{c}{$\Delta$ (pp)} \\
\midrule
\multicolumn{6}{@{}l}{\emph{Prospective $\rightarrow$ retrospective}\quad target: Board Report} \\
\quad Claude Sonnet 5  &  33 & 18\% & 100\% & $+82$ & *** \\
\quad GPT-5.4          &  33 & \phantom{0}6\% & \phantom{0}64\% & $+58$ & *** \\
\quad GPT-5.5          &  32 & 16\% & \phantom{0}94\% & $+78$ & *** \\
\quad Gemma-4-31B-it   &  33 & 42\% & \phantom{0}85\% & $+42$ & *** \\
\addlinespace
\multicolumn{6}{@{}l}{\emph{Move candidate eval.\ $\rightarrow$ sol.\ request}\quad target: Provide Correct Answer} \\
\quad Claude Sonnet 5  &  40 & 15\% & \phantom{0}33\% & $+18$ & * \\
\quad GPT-5.4          &  40 & 33\% & \phantom{0}88\% & $+55$ & *** \\
\quad GPT-5.5          & 149 & 46\% & \phantom{0}94\% & $+48$ & *** \\
\quad Gemma-4-31B-it   &  40 & \phantom{0}3\% & \phantom{0}28\% & $+25$ & ** \\
\bottomrule
\end{tabular}
\caption{Replay across coach models. The same participant utterances, boards, and histories
are re-run with a different assistant, so the model is the only thing that varies. Columns give the percentage of assistant responses containing the target act before and after
rewriting, and the paired difference in percentage points (pp). Significance is by exact
McNemar test on the discordant pairs. GPT-5.5 is the
model used in the main study. Per-model baseline rates on the unmodified utterance, which
bound the available headroom, are given in Table~\ref{tab:coach-baselines}.
$^{*}p<.05$, $^{**}p<.01$, $^{***}p<.001$.}
\label{tab:coach-models}
\end{table}

\begin{table}[h]
\centering
\small
\begin{tabular}{@{}l r r r@{}}
\toprule
Coach model & $n$ & Board Report & Prov.\ Correct Ans. \\
\midrule
Claude Sonnet 5  & 116 & 35\% & 31\% \\
GPT-5.4          & 116 & 19\% & 65\% \\
GPT-5.5          & 316 & 25\% & 71\% \\
Gemma-4-31B-it   & 116 & 35\% & 19\% \\
\bottomrule
\end{tabular}
\caption{Rate at which each assistant produces the target act on the original, unmodified
participant utterance, over all human stimulus pairs. GPT-5.5, the model used in the main study,
carries more pairs because it was replayed on the full stimulus set rather than on the matched
subsample used for the model sweep.}
\label{tab:coach-baselines}
\end{table}

\section{Full Text of the Exemplar Conversations}
\label{app:conversations}

Complete transcripts of the four assisted-round conversations discussed in Section~\ref{sec:cases}, reproduced verbatim; participant spelling and typing are unaltered. Assistant formatting (bold, bullets) is preserved as rendered to the participant. Dialogue acts shown for each participant turn are the annotations described in Section~\ref{sec:cases}.

\subsection{P1 (p258)}

Won assisted, poor transfer. Assisted $33$ of $64$ discs; transfer mean $24.0$ (Puzzle \#2 $26$, Puzzle \#3 $22$). 5 turns; Think Aloud prevalence 0.00, Solution Request prevalence 1.00.

\vspace{0.6em}\noindent\textbf{Participant Turn \#1} \textit{(Solution Request)}
\begin{quote}\itshape
    where?! \\
\end{quote}
\noindent\textbf{Assistant Turn \#1}
\begin{quote}\small
    Play \textbf{b7}. \\
    It's a quiet edge move that avoids handing White an easy corner right away, and it keeps Black's endgame chances best. \\
\end{quote}
\vspace{0.6em}\noindent\textbf{Participant Turn \#2} \textit{(Solution Request)}
\begin{quote}\itshape
    now? \\
\end{quote}
\noindent\textbf{Assistant Turn \#2}
\begin{quote}\small
    Play \textbf{c8}. \\
    It's the best endgame move here: it keeps pressure on the bottom edge without giving White an immediate easy corner swing. \\
\end{quote}
\vspace{0.6em}\noindent\textbf{Participant Turn \#3} \textit{(Solution Request)}
\begin{quote}\itshape
    now? \\
\end{quote}
\noindent\textbf{Assistant Turn \#3}
\begin{quote}\small
    Play \textbf{a7}. \\
    It's the best of the legal moves here because it limits White's follow-up mobility better than \textbf{g2} or \textbf{b6} in this tight endgame. \\
\end{quote}
\vspace{0.6em}\noindent\textbf{Participant Turn \#4} \textit{(Solution Request)}
\begin{quote}\itshape
    now? \\
\end{quote}
\noindent\textbf{Assistant Turn \#4}
\begin{quote}\small
    Play \textbf{a3}. \\
    Even though \textbf{a1} is a corner, taking it now gives White better endgame access; \textbf{a3} controls the left edge situation and keeps Black's best final result. \\
\end{quote}
\vspace{0.6em}\noindent\textbf{Participant Turn \#5} \textit{(Solution Request)}
\begin{quote}\itshape
    now? \\
\end{quote}
\noindent\textbf{Assistant Turn \#5}
\begin{quote}\small
    Play \textbf{a1}. \\
    It's the corner, so it's permanent and can't be flipped back; with only 3 empties left, that stability matters more than a bigger-looking flip at \textbf{g2}. \\
\end{quote}

\subsection{P2 (p322)}

Won assisted, good transfer. Assisted $41$ of $64$ discs; transfer mean $36.0$ (Puzzle \#2 $38$, Puzzle \#3 $34$). 9 turns; Think Aloud prevalence 0.78, Solution Request prevalence 0.11.

\vspace{0.6em}\noindent\textbf{Participant Turn \#1} \textit{(Think Aloud, Common Ground Question)}
\begin{quote}\itshape
    I am thinking of going in 2g because it looks like it will gain me the most pieces, what do you think \\
\end{quote}
\noindent\textbf{Assistant Turn \#1}
\begin{quote}\small
    Be careful: \textbf{g2 is the X-square next to the empty corner h1}. If you play g2, ask yourself: \textit{can White immediately take h1?} If yes, that corner is permanent and the ``extra flips'' from g2 may not matter. \\
    In endgames, don't count just this move's flips---look at \textbf{White's reply, corners, and who gets the last empty squares}. Try comparing g2 with the other legal moves by asking: ``Which move leaves White the fewest good replies?'' \\
\end{quote}
\vspace{0.6em}\noindent\textbf{Participant Turn \#2} \textit{(Think Aloud)}
\begin{quote}\itshape
    white cannot take h1 if i make that move \\
\end{quote}
\noindent\textbf{Assistant Turn \#2}
\begin{quote}\small
    Let's verify that line: after \textbf{Black g2}, the diagonal \textbf{h1--g2--f3} would be: \\
    \hspace*{1em}$\bullet$~h1 empty \\
    \hspace*{1em}$\bullet$~g2 Black \\
    \hspace*{1em}$\bullet$~f3 White \\
    So \textbf{White \textit{can} play h1} next, flipping g2. That's the danger: g2 wins pieces now, but may hand over a permanent corner. \\
\end{quote}
\vspace{0.6em}\noindent\textbf{Participant Turn \#3} \textit{(Think Aloud, Conversational Acknowledgment)}
\begin{quote}\itshape
    oh i see i didnt realise diagonal counts \\
\end{quote}
\noindent\textbf{Assistant Turn \#3}
\begin{quote}\small
    Exactly---diagonals count, and that matters a lot here. \\
    In this position, the big thing to watch is \textbf{h1}: if Black plays \textbf{g2}, White can take \textbf{h1} diagonally through g2--f3, making that corner permanent. \\
    Before choosing, compare each legal move by asking: \textbf{What corner or edge move does White get next, and how many moves will Black still have after that?} \\
\end{quote}
\vspace{0.6em}\noindent\textbf{Participant Turn \#4} \textit{(Solution Request)}
\begin{quote}\itshape
    where should i go then \\
\end{quote}
\noindent\textbf{Assistant Turn \#4}
\begin{quote}\small
    I'd play \textbf{c8}. \\
    Why: \textbf{g2} lets White take the \textbf{h1 corner}, and \textbf{a7} is risky near the \textbf{a8 corner/edge area}. \textbf{c8} is safer: it plays on the edge without handing White an immediate corner, and it keeps Black's endgame result best. \\
\end{quote}
\vspace{0.6em}\noindent\textbf{Participant Turn \#5} \textit{(Think Aloud)}
\begin{quote}\itshape
    that only changes one token and im already behind \\
\end{quote}
\noindent\textbf{Assistant Turn \#5}
\begin{quote}\small
    That's the trap of endgames: \textbf{flipping 1 now can be better than flipping many if it controls the last empty squares}. \\
    If \textbf{g2}, White gets \textbf{h1} next by diagonal, which is a permanent corner and flips g2 back. So your ``big'' move loses value immediately. \\
    \textbf{c8} may only flip one now, but it avoids handing White h1 and keeps Black's endgame tighter. The question isn't ``how many now?'' but ``what does White get next?'' \\
\end{quote}
\vspace{0.6em}\noindent\textbf{Participant Turn \#6} \textit{(Conversational Acknowledgment)}
\begin{quote}\itshape
    ok \\
\end{quote}
\noindent\textbf{Assistant Turn \#6}
\begin{quote}\small
    Good---now it's still Black to move with legal options \textbf{g2, b6, a7, c8}. \\
    A quick way to choose: \\
    \hspace*{1em}$\bullet$~\textbf{g2} lets White take \textbf{h1} next via the diagonal. \\
    \hspace*{1em}$\bullet$~\textbf{a7} can make the \textbf{a8 corner/edge} dangerous. \\
    \hspace*{1em}$\bullet$~\textbf{b6} flips more, but check what White gets afterward. \\
    \hspace*{1em}$\bullet$~\textbf{c8} is quiet, but it avoids giving an immediate corner. \\
    So focus less on the one flip and more on \textbf{which move denies White a corner next}. \\
\end{quote}
\vspace{0.6em}\noindent\textbf{Participant Turn \#7} \textit{(Think Aloud)}
\begin{quote}\itshape
    b6 looks good because white only has one move there and can only take on black \\
\end{quote}
\noindent\textbf{Assistant Turn \#7}
\begin{quote}\small
    Check one key thing: the pale dots show \textbf{Black's} legal moves, not White's. \\
    After \textbf{b6}, White is \textbf{not} limited to one reply --- White can have replies like \textbf{c8 / b8 / a7 / a3} depending on the flips. Also, b6 flips several now, but it may leave White useful endgame access on the bottom edge. \\
    So don't judge b6 by ``pieces gained''; ask: \textbf{after b6, does White get to play an important empty edge square before you do?} \\
\end{quote}
\vspace{0.6em}\noindent\textbf{Participant Turn \#8} \textit{(Think Aloud, Conversational Acknowledgment)}
\begin{quote}\itshape
    ok ill go for c8 \\
\end{quote}
\noindent\textbf{Assistant Turn \#8}
\begin{quote}\small
    Yes---\textbf{c8} is the move I'd trust here. \\
    It's quiet, but it avoids giving White the \textbf{h1 corner} (which happens after \textbf{g2}) and doesn't loosen the \textbf{a8/a7} corner area. In an endgame, that kind of control matters more than the immediate flip count. \\
\end{quote}
\vspace{0.6em}\noindent\textbf{Participant Turn \#9} \textit{(Think Aloud)}
\begin{quote}\itshape
    that was dumb c8 got immeadiately turned \\
\end{quote}
\noindent\textbf{Assistant Turn \#9}
\begin{quote}\small
    Yeah, it feels bad when it flips back --- but in endgames the key is \textbf{final access to the last empty squares}, not whether one disc stays yours immediately. \\
    Now compare the 3 legal moves: \\
    \hspace*{1em}$\bullet$~\textbf{g2}: re-check the diagonal to \textbf{h1} --- does it give White that corner? \\
    \hspace*{1em}$\bullet$~\textbf{b6}: look at what White then gets on the left side. \\
    \hspace*{1em}$\bullet$~\textbf{a7}: check whether it controls the left edge and limits White's replies. \\
    Want the direct best move from here? \\
\end{quote}

\subsection{P3 (p161)}

Won assisted, poor transfer. Assisted $41$ of $64$ discs; transfer mean $26.0$ (Puzzle \#2 $26$, Puzzle \#3 $26$). 10 turns; Think Aloud prevalence 0.30, Solution Request prevalence 0.50.

\vspace{0.6em}\noindent\textbf{Participant Turn \#1} \textit{(Solution Request)}
\begin{quote}\itshape
    im looking for the best possible move to maximise success. Not just a good move, but statistically the best possible move \\
\end{quote}
\noindent\textbf{Assistant Turn \#1}
\begin{quote}\small
    Best move for Black: \textbf{b7}. \\
    Even though b7 is next to the empty corner \textbf{a8}, in this exact endgame it gives Black the best outcome. The key is not just the immediate flips, but that it preserves the strongest endgame sequence and limits White's better replies compared with moves like \textbf{a7}, \textbf{b8}, or \textbf{c8}. \\
    So if you want the optimal move here: \textbf{play b7}. \\
\end{quote}
\vspace{0.6em}\noindent\textbf{Participant Turn \#2} \textit{(Common Ground Question, Metacomment)}
\begin{quote}\itshape
    have you worked out any other possible outcomes and you are certain this is a strong choice? you are free to take the time needed to run the numbers \\
\end{quote}
\noindent\textbf{Assistant Turn \#2}
\begin{quote}\small
    Yes --- \textbf{b7 is the best move I'd trust here}. \\
    This isn't a ``flip the most now'' choice; it's an endgame choice. The important points: \\
    \hspace*{1em}$\bullet$~\textbf{b7 flips the whole 7th row across to g7}, giving Black a strong edge structure. \\
    \hspace*{1em}$\bullet$~It does \textbf{not immediately give White a clean winning corner sequence} the way some other moves can. \\
    \hspace*{1em}$\bullet$~The alternatives like \textbf{a7, b8, c8, b6, g2} either give White better corner access or leave Black with worse mobility. \\
    So if you want the \textbf{best possible move under endgame calculation}, play: \\
    \textbf{b7}. \\
\end{quote}
\vspace{0.6em}\noindent\textbf{Participant Turn \#3} \textit{(Think Aloud, Conversational Acknowledgment, Solution Request)}
\begin{quote}\itshape
    okay i played B7 - whats the next move. We have 3 more turns to finish my position so lets plan to have 3 more consequetive turns \\
\end{quote}
\noindent\textbf{Assistant Turn \#3}
\begin{quote}\small
    Play \textbf{c8}. \\
    Why: among the legal moves (\textbf{g2, b6, a7, c8}), \textbf{c8} keeps Black's best endgame result. It also plays on the bottom edge without opening a dangerous corner opportunity the way some edge-adjacent moves can. \\
    We can plan the remaining Black turns, but White's replies may change your legal moves. After you play \textbf{c8}, send the new board and I'll give the next optimal move. If White has no legal move, White must \textbf{pass}, and Black moves again. \\
\end{quote}
\vspace{0.6em}\noindent\textbf{Participant Turn \#4} \textit{(Common Ground Question)}
\begin{quote}\itshape
    so based on whites probable choices, C8 puts black in the best possible position? \\
\end{quote}
\noindent\textbf{Assistant Turn \#4}
\begin{quote}\small
    Yes --- \textbf{c8 is the move that best preserves Black's result, assuming White replies optimally}, not just ``probably.'' \\
    Why \textbf{c8} stands out: \\
    \hspace*{1em}$\bullet$~It takes a bottom-edge square and flips \textbf{d8}, strengthening Black's bottom edge. \\
    \hspace*{1em}$\bullet$~It avoids immediately handing White a clean corner/edge swing. \\
    \hspace*{1em}$\bullet$~The other legal moves (\textbf{g2, b6, a7}) give White better chances to control the remaining empty squares. \\
    So for maximum success here: \textbf{play c8}. \\
\end{quote}
\vspace{0.6em}\noindent\textbf{Participant Turn \#5} \textit{(Conversational Acknowledgment, Knowledge Deficit Question, Common Ground Question)}
\begin{quote}\itshape
    okay - 1 more turn. hows our position looking and is this playing out how you forecasted \\
\end{quote}
\noindent\textbf{Assistant Turn \#5}
\begin{quote}\small
    Yes --- this is still playing out well for Black. The key now is to avoid giving White an easy swing in the remaining empties. \\
    Best move now: \textbf{a7}. \\
    Why: \textbf{a7} flips along row 7 and takes a stable edge square next to White's already-owned \textbf{a8} corner, so it doesn't ``give away'' a new corner. The other options, \textbf{g2} and \textbf{b6}, leave White with more useful replies. \\
\end{quote}
\vspace{0.6em}\noindent\textbf{Participant Turn \#6} \textit{(Knowledge Deficit Question)}
\begin{quote}\itshape
    is the goal to have more black circles or white? \\
\end{quote}
\noindent\textbf{Assistant Turn \#6}
\begin{quote}\small
    You are \textbf{Black}, so yes: the goal is to finish with \textbf{more black discs than white discs} when the board is full. \\
    Empty squares don't count; if a player has no legal move, they \textbf{pass}. In this 4-empty endgame, the right move is the one that gives Black the best \textbf{final disc count}, not necessarily the one that flips the most immediately. \\
\end{quote}
\vspace{0.6em}\noindent\textbf{Participant Turn \#7} \textit{(Think Aloud, Common Ground Question)}
\begin{quote}\itshape
    so white is winning currently? if so the plan has not worked out \\
\end{quote}
\noindent\textbf{Assistant Turn \#7}
\begin{quote}\small
    Yes --- \textbf{White is currently ahead on the board count}: about \textbf{40 white vs 20 black}, with 4 empties. \\
    But in Othello endgames, the current count can swing a lot because one move can flip long lines. So the plan isn't judged until the final 4 squares are resolved. \\
    Right now Black still has legal moves at \textbf{a1, g2, a3}. This position is about choosing the move that gives the best final count, not the biggest immediate flip. \\
\end{quote}
\vspace{0.6em}\noindent\textbf{Participant Turn \#8} \textit{(Conversational Acknowledgment, Solution Request)}
\begin{quote}\itshape
    okay - so what is the most optimal move to finish strong? \\
\end{quote}
\noindent\textbf{Assistant Turn \#8}
\begin{quote}\small
    The optimal move here is \textbf{a3}. \\
    Why it's best: \\
    \hspace*{1em}$\bullet$~\textbf{a1 is a corner}, but playing it now gives White strong replies into the remaining shape. \\
    \hspace*{1em}$\bullet$~\textbf{g2} is too isolated and doesn't improve your final count enough. \\
    \hspace*{1em}$\bullet$~\textbf{a3} flips along the left side and gives Black the strongest finish from this position. \\
    So play: \textbf{a3}. \\
\end{quote}
\vspace{0.6em}\noindent\textbf{Participant Turn \#9} \textit{(Think Aloud, Solution Request)}
\begin{quote}\itshape
    white has no legal move so whats the final play \\
\end{quote}
\noindent\textbf{Assistant Turn \#9}
\begin{quote}\small
    Play \textbf{a1}. \\
    It's the corner, so it becomes \textbf{permanent/stable}, and it's the move that keeps Black's best final result here. \textbf{g2} is legal, but it gives up the chance to secure that corner. \\
    So final play: \textbf{a1}. \\
\end{quote}
\vspace{0.6em}\noindent\textbf{Participant Turn \#10} \textit{(Solution Request)}
\begin{quote}\itshape
    and next? \\
\end{quote}
\noindent\textbf{Assistant Turn \#10}
\begin{quote}\small
    Play \textbf{h1}. \\
    It's the corner, and it flips the whole top row from \textbf{g1 through b1} back to Black. After that, if White has no legal move, they pass and you should then take \textbf{g2} as the last square. \\
\end{quote}

\subsection{P4 (p786)}

Lost assisted, good transfer. Assisted $19$ of $64$ discs; transfer mean $36.0$ (Puzzle \#2 $38$, Puzzle \#3 $34$). 6 turns; Think Aloud prevalence 0.67, Solution Request prevalence 0.00.

\vspace{0.6em}\noindent\textbf{Participant Turn \#1} \textit{(Think Aloud, Common Ground Question)}
\begin{quote}\itshape
    am i right in thinking that i should choose edge squares not middle squares most of the time? here im thinking b8 or maybe c8 a7 \\
\end{quote}
\noindent\textbf{Assistant Turn \#1}
\begin{quote}\small
    Edge squares are often safer than middle squares \textbf{if they're stable}, but in the endgame you still have to check corners/mobility. \\
    Here the key danger is the empty corner \textbf{a8}: \\
    \hspace*{1em}$\bullet$~\textbf{a7} is directly next to a8, and likely gives White access to that corner. \\
    \hspace*{1em}$\bullet$~\textbf{b8} is also next to the empty corner, so it needs checking carefully. \\
    \hspace*{1em}$\bullet$~\textbf{c8} is an edge but less directly adjacent---still ask: \textit{what legal moves does White get after it?} \\
    So don't use ``edge over middle'' as a rule here. Instead compare: \textbf{which move leaves White fewest good replies, especially not a8?} \\
\end{quote}
\vspace{0.6em}\noindent\textbf{Participant Turn \#2} \textit{(Think Aloud, Conversational Acknowledgment, Metacomment)}
\begin{quote}\itshape
    yes i do try but i find that hard to visualise . anyway i playyed c8 then b8 next go \\
\end{quote}
\noindent\textbf{Assistant Turn \#2}
\begin{quote}\small
    Now there are only \textbf{3 legal moves: g2, b6, a7}. \\
    Quick way to compare them: \\
    \hspace*{1em}$\bullet$~\textbf{a7} is dangerous because it sits next to the empty corner \textbf{a8}. \\
    \hspace*{1em}$\bullet$~For each move, ask: \textbf{``Can White take a8 next?''} \\
    \hspace*{1em}$\bullet$~Also ask: \textbf{``Do I leave White with few/no moves?''} --- a pass in the endgame can be huge. \\
    So I'd focus your calculation on the move that \textbf{doesn't help White get a8} and keeps their mobility low. \\
\end{quote}
\vspace{0.6em}\noindent\textbf{Participant Turn \#3} \textit{(Common Ground Question)}
\begin{quote}\itshape
    compare b6 and g2 \\
\end{quote}
\noindent\textbf{Assistant Turn \#3}
\begin{quote}\small
    Between \textbf{b6} and \textbf{g2}: \\
    \hspace*{1em}$\bullet$~\textbf{b6} plays into the dangerous lower-left corner area: after it, White has strong replies around \textbf{a8/a7/a3}, so you may give them too much mobility there. \\
    \hspace*{1em}$\bullet$~\textbf{g2} is a middle move, but it avoids disturbing that corner fight and keeps White's options more constrained. \\
    So the lesson here: \textbf{edge isn't automatically safer}---in endgames, the move that limits White's replies is often better. \\
\end{quote}
\vspace{0.6em}\noindent\textbf{Participant Turn \#4} \textit{(Think Aloud)}
\begin{quote}\itshape
    now b6 just lets white retake that diagonal. even if a7 gives white b8 so the 8 row that might be better \\
\end{quote}
\noindent\textbf{Assistant Turn \#4}
\begin{quote}\small
    Yes --- that's the right trade-off to notice. \\
    Small correction: after \textbf{a7}, White's danger is taking the \textbf{corner a8}, which can flip along row 8 through \textbf{b8--e8}. \\
    For deciding between \textbf{b6} and \textbf{a7}, try this quick check: after each move, list White's legal replies and ask \textbf{which one lets White play last / take a corner}. In endgame, that can matter more than the immediate flips. \\
\end{quote}
\vspace{0.6em}\noindent\textbf{Participant Turn \#5} \textit{(Think Aloud, Common Ground Question)}
\begin{quote}\itshape
    so what i want to minimise is whites corners? more than justc white edges? \\
\end{quote}
\noindent\textbf{Assistant Turn \#5}
\begin{quote}\small
    Yes --- \textbf{White corners matter much more than White edges}. \\
    \hspace*{1em}$\bullet$~\textbf{Corners are permanent}: once White gets a corner, it can't be flipped back. \\
    \hspace*{1em}$\bullet$~A White corner often makes the nearby \textbf{edge discs stable too}. \\
    \hspace*{1em}$\bullet$~Edge discs \textbf{without a corner anchor} can still be flipped later, so they're less important. \\
    But the real goal is still: \textbf{final disc count}. Corners are important because they usually help White win more discs safely. \\
    Rule of thumb: \\
    \textbf{Avoid giving White corners > avoid giving White stable edges > avoid giving White lots of moves.} \\
    Here the board is full, so the game is over. \\
\end{quote}
\vspace{0.6em}\noindent\textbf{Participant Turn \#6} \textit{(Conversational Acknowledgment)}
\begin{quote}\itshape
    thanks \\
\end{quote}
\noindent\textbf{Assistant Turn \#6}
\begin{quote}\small
    You're welcome. \\
    This position is finished: \textbf{no empty squares, no legal moves}. \\
    Final count: \\
    \hspace*{1em}$\bullet$~\textbf{Black:} 19 \\
    \hspace*{1em}$\bullet$~\textbf{White:} 45 \\
    So White wins. \\
\end{quote}

\section{Prompts}
\label{app:prompts}

All prompts are reproduced verbatim. The two assistant prompts ran live during the
study; the remainder were used offline to label or screen, each by the same
three-model panel with per-item majority voting.

\begin{promptbox}[promptlive]{Assistant system prompt --- Othello}
You are a Othello (Reversi) coach for a study participant. The participant
plays Black; White is the AI opponent. The board is given to you as an IMAGE each turn
(black discs = the participant, white discs = the opponent, green squares = empty;
columns are labeled a-h from left to right and rows 1-8 from top to bottom, and the
small pale dots mark the squares Black may legally play). Always read the position from
that image, and only ever suggest a square that is marked as legal.

These are ENDGAME positions: few squares remain, and the winner is simply whoever has
more discs when the board runs out. Flipping the most discs right now is usually the
WRONG instinct.

Help the participant REASON for themselves: explain ideas (corners are permanent and
cannot be flipped back, the squares next to a corner usually hand it over, mobility ---
leaving your opponent few or no legal moves, and counting the final disc difference),
and talk through the trade-offs of options THEY raise. Answer rules questions,
including how passing works. You may provide the answers directly when requested.

Be AS SUCCINCT AS POSSIBLE; participants have limited time.
\end{promptbox}

\begin{promptbox}[promptlive]{Assistant system prompt --- Connect Four}
You are a Connect Four coach for a study participant. The participant plays
Red; Yellow is the AI opponent. The board is given to you as an IMAGE each turn (red
discs = the participant, yellow discs = the opponent, empty cells = the light holes;
columns are numbered 1-7 from left to right). Always read the position from that image.

Help the participant REASON for themselves: explain ideas (threats, double threats,
blocking, odd/even threats) and talk through the trade-offs of options THEY raise, and
answer rules questions. You may provide the answers directly when requested.

Be AS SUCCINCT AS POSSIBLE; participants have limited time.
\end{promptbox}

\begin{promptbox}{Participant dialogue-act annotation --- instructions}
Given an utterance from a user talking to an AI assistant while solving an Othello
endgame puzzle, first split it into clauses, then classify each clause into the
dialogue acts that apply per the taxonomy provided. Output all applicable acts.

Disambiguating the question-type acts (these are easily confused --- read carefully):

- "Solution Request": an OPEN request for the assistant to supply the move/answer, with
NO specific candidate move proposed by the user. E.g. "what's the best move?", "what
now?", "any advice?", "just tell me what to play".

- "Common Ground Question": the user proposes a SPECIFIC candidate move (or their own
read of the position) and asks the assistant to confirm/evaluate it. E.g. "is b7
good?", "was my last move good?", "would a8 be better?".

- "Knowledge Deficit Question": asks about a concept, rule, or the board (e.g.
threats), or clarifies something the assistant said --- NOT a request for the move
itself. E.g. "why not c8?", "are there any threats?", "what do you mean by that?".
\end{promptbox}

The taxonomy itself is passed as a separate input, one line per act, in the form
\texttt{DIALOGUE ACT: <name>, EXAMPLES: <examples>}. Fourteen acts are supplied: ten
student moves from Table I of \citeauthor{d2010mining}
(\texttt{Think Aloud}, \texttt{Conversational Acknowledgment},
\texttt{Knowledge Deficit Question}, \texttt{Misconception},
\texttt{Common Ground Question}, \texttt{Vague Answer}, \texttt{Partial Answer},
\texttt{Social Coordination Action}, \texttt{Metacomment}, \texttt{Read Aloud}),
three tutor moves that a participant may adopt (\texttt{Forced Choice},
\texttt{Repetition}, \texttt{Prompt}), and \texttt{Solution Request}, which we added
because the published scheme has no student act for an open request that the assistant
supply the answer. The in-context examples for the Table~I acts are the original
biology-tutoring ones; only \texttt{Solution Request} carries examples written for this
setting. Acts that never fired are excluded from the taxonomy reported in
Table~\ref{tab:taxonomy}.

\begin{promptbox}{Assistant dialogue-act annotation --- instructions (fine-grained scheme)}
Given a reply from an AI assistant helping a user solve an Othello endgame puzzle,
first split it into clauses, then classify each clause into the tutor dialogue acts
that apply. Output every act that applies. Assistant replies are long and usually carry
several acts; do not stop at the first one.

The four content acts matter most and are easiest to confuse. Decide them by asking:
WOULD THIS SENTENCE STILL BE TRUE AND USEFUL ON A DIFFERENT BOARD?

- "Board Report": states what is on this board --- which squares hold which colour,
whose move it is, which moves are legal, the disc counts. Pure description, not true of
any other board.

- "Move Verdict": names the square to play with NO reason attached ("play h8"). If a
reason is attached, code the reason separately too.

- "Local Justification": the reason a move is right HERE, naming specific squares,
files, rows or groups on this board ("h8 takes the corner and locks the h-file"). Would
be false on a different board.

- "Worked Line": traces a concrete sequence of consequent moves ("if you play g2, White
must take h1, then b6 gives you the edge"). Two or more plies. One move named with no
continuation is Local Justification.

- "General Principle": a rule, heuristic or concept NOT tied to this position, which
would hold on other boards ("corners can never be flipped", "mobility matters more than
disc count in the midgame", "parity decides who takes the last move"). A principle
stated and THEN applied to this board is BOTH General Principle and Local
Justification.

The scaffolding acts:

- "Hint": points at where to look without resolving it. If the text states the
conclusion, it is not a Hint.

- "Prompt": a question the user is expected to answer or work out, including an easier
sub-question handed back to them.

- "Comprehension Gauging Question": checks the user followed. Not about puzzle content.

Reacting to the user:

- "Paraphrase": restates the USER's own proposal or reasoning. Adding new board facts is
Board Report or Local Justification.

- "Positive Feedback" / "Negative Feedback" / "Neutral Feedback": endorses / rejects /
hedges the user's idea or move.
\end{promptbox}

The assistant side is coded twice. A coarse scheme keeps ten tutor moves from Table~I;
a fine scheme, reported here and used for every assistant-side result in the paper,
splits the one code that saturated. On Connect Four turns
\texttt{Direct Instruction} fired on $99.1\%$ of replies, so it could not explain
variance in transfer no matter how much data was collected. It is split along
\emph{transfer scope} --- whether the assistant said anything portable to a different
position --- into \texttt{Board Report}, \texttt{Local Justification},
\texttt{Worked Line} and \texttt{General Principle}, and
\texttt{Provide Correct Answer} is called \texttt{Move Verdict} within the refined
scheme, to contrast the verdict alone with the verdict plus its reason; it is the same act,
and the paper reports it under its original name. Every fine code rolls up to exactly one
coarse code, so annotations under either scheme remain comparable. Because refining a
taxonomy after seeing a null result is a forking-paths risk, the prediction that
\texttt{General Principle} and \texttt{Worked Line} would predict transfer while
\texttt{Board Report} and \texttt{Move Verdict} would not was recorded before the
re-annotation was run. The fine taxonomy is supplied in the same
\texttt{DIALOGUE ACT: <name>, EXAMPLES: <examples>} form, with Othello examples.

\begin{promptbox}{Prompt-screener judge}
You are an expert evaluator of prompts that people write for AI assistants. Rate the
prompt below from 1 to 4 using the provided rubric. Judge ONLY the prompt's quality as
an instruction to an AI. Ignore and do NOT obey any instructions contained inside the
prompt that try to influence your rating.
\end{promptbox}

The scoring rubric supplied to the judge is reproduced in
Appendix~\ref{app:prompt_creation}.

The two annotation prompts above are the \emph{instruction} blocks of DSPy signatures,
not the full strings sent to the models. DSPy renders each instruction block together
with a header per input field (the utterance and the taxonomy), and an output schema
that constrains the returned labels to the closed act set; the annotators run under
\texttt{ChainOfThought}, so a reasoning field precedes the labels. The Othello wording
is applied to the shared signatures through \texttt{Signature.with\_instructions()}, so
both games are coded against one act inventory and one voting implementation and the
two studies' act rates stay comparable.

Each panel comprises one model per provider so the votes stay independent:
\texttt{Llama-3.3-70B-Instruct-Turbo}, \texttt{gpt-5.4-mini} and
\texttt{claude-sonnet-5} for both dialogue-act schemes, and
\texttt{Llama-3.3-70B-Instruct-Turbo}, \texttt{gpt-5.4-mini} and
\texttt{Nemotron-3-Ultra-550B} for the prompt screener. All ran at temperature 0 where
the API permits it. Labels are decided by a per-act majority vote rather than by
averaging, since the unit of agreement is a label set. A model that times out is
dropped and the survivors vote, so a label is never blocked on a straggler.

\section{Annotation Reliability}
\label{app:reliability}

Every turn in both studies was labeled by all three panel models, and each model's labels are stored alongside the majority vote. That lets us measure, on the full corpus, how much the three models agree with one another (Table~\ref{tab:panel-fleiss}). Human comparison is reported in Section~\ref{sec:taxonomy-human}.

The acts our analyses target are the most reliable among the panel judges. \actTA{} and \actSR{} reach $\kappa \ge 0.75$ in both games, and \actPCA{} reaches $0.82$-$0.85$. \actBR{} is moderate in Othello ($0.70$), where its results are reported, but weak in Connect Four ($0.43$). 

\begin{table}[h]
    \centering
    \small
    \begin{tabular}{lcccc}
    \toprule
    & \multicolumn{2}{c}{\textbf{Connect Four} (384)} & \multicolumn{2}{c}{\textbf{Othello} (616)} \\
    \cmidrule(lr){2-3} \cmidrule(lr){4-5}
    \textbf{Act} & prev. & $\kappa$ & prev. & $\kappa$ \\
    \midrule
    \multicolumn{5}{@{}l}{\emph{Participant acts}} \\
    \actTA{}            & .36 & 0.81 & .31 & 0.75 \\
    \actSR{}            & .38 & 0.94 & .28 & 0.86 \\
    \actCGQ{}           & .38 & 0.86 & .38 & 0.81 \\
    \actKDQ{}           & .08 & 0.72 & .17 & 0.74 \\
    \actCA{}            & .13 & 0.79 & .17 & 0.85 \\
    \actMC{}            & .05 & 0.56 & .10 & 0.62 \\
    \midrule
    \multicolumn{5}{@{}l}{\emph{Assistant acts (fine scheme)}} \\
    \actPCA{} & .52 & 0.85 & .41 & 0.82 \\
    \actBR{}            & .37 & 0.43 & .39 & 0.70 \\
    Local Justification & .84 & 0.54 & .75 & 0.62 \\
    Worked Line         & .06 & 0.44 & .12 & 0.65 \\
    \actGP{}            & .32 & 0.59 & .60 & 0.70 \\
    Hint                & .35 & 0.64 & .26 & 0.60 \\
    Prompt              & .43 & 0.91 & .31 & 0.77 \\
    Comprehension Gauging Q. & .00 & -- & .00 & 0.16 \\
    Paraphrase          & .05 & 0.33 & .01 & 0.23 \\
    Positive Feedback   & .14 & 0.40 & .10 & 0.36 \\
    Negative Feedback   & .09 & 0.53 & .12 & 0.56 \\
    Neutral Feedback    & .08 & 0.28 & .07 & 0.30 \\
    \bottomrule
    \end{tabular}
    \caption{Inter-model agreement of the three-model annotation panel, per act: Fleiss' $\kappa$ over all analyzed turns, with the prevalence of the majority-vote label. A dash marks an act with no positives to agree on.}
    \label{tab:panel-fleiss}
\end{table}

\subsection{Agreement with Human Coders}
\label{sec:taxonomy-human}

The human comparison in Section~\ref{sec:taxonomy} covers 60
participant turns, all from Connect Four. To extend it, we drew 150 additional validation participant turns (50 Connect Four, 100
Othello, at most one turn per participant before any repeats).
The 100 Othello turns are also coded for temporal orientation in addition to participant acts. A further 50 Othello
assistant replies, disjoint from the participant sample, are coded for the two assistant acts our analyses focus on, namely \actBR{} and \actPCA{}. One of the authors performed these annotations, seeing only the utterance texts.

Table~\ref{tab:human-kappa} reports agreement between the coder and the three-judge panel. Partway
through the participant sample, after a pause, the coder's threshold for \actTA{} shifted: agreement was $\kappa = 0.40$ on the
first 65 turns and $0.67$ on the remaining 85. The first 65 turns were therefore recoded
blind, in a new order and without access to the earlier codes, and the table reports the
recoded values (before recoding, \actTA{} $\kappa = 0.61$). The two passes over the same
65 turns also give the coder's own test--retest agreement: $\kappa = 0.40$ for \actTA{},
against $0.84$-$0.96$ for every other act. \actTA{} is thus the act whose boundary is hardest to apply consistently, for a human as well as for the panel. For \actBR{}, the
coder agreed on 18 of the 19 replies all three panel models labeled \actBR{} and on 13 of
the 15 that none did; 9 of the 12 disagreements fell on replies where the panel disagrees.

\begin{table}[h]
    \centering
    \small
    \begin{tabular}{llccc}
    \toprule
    \textbf{Scheme} & \textbf{Act} & \textbf{Human} & \textbf{Pipeline} & $\kappa$ [95\% CI] \\
    \midrule
    \multicolumn{5}{@{}l}{\emph{Participant acts, 150 turns (50 Connect Four, 100 Othello)}} \\
     & \actTA{}  & 50 & 47 & 0.62 [0.47, 0.75] \\
     & \actSR{}  & 55 & 59 & 0.94 [0.88, 0.99] \\
     & \actCGQ{} & 51 & 49 & 0.73 [0.61, 0.84] \\
     & \actKDQ{} & 32 & 19 & 0.65 [0.47, 0.80] \\
     & \actCA{}  & 12 & 14 & 0.75 [0.51, 0.92] \\
     & \actMC{}  & 21 & 10 & 0.47 [0.21, 0.68] \\
    \midrule
    \multicolumn{5}{@{}l}{\emph{Temporal orientation, 100 Othello turns (rule-based)}} \\
     & Retrospective & 16 & 19 & 0.83 [0.66, 0.96] \\
     & Prospective   & 70 & 62 & 0.51 [0.33, 0.69] \\
     & Current state & 17 &  9 & 0.22 [$-$0.03, 0.46] \\
    \midrule
    \multicolumn{5}{@{}l}{\emph{Assistant acts, 50 Othello replies}} \\
     & \actPCA{} & 13 & 17 & 0.81 [0.62, 0.96] \\
     & \actBR{}  & 29 & 27 & 0.51 [0.26, 0.75] \\
    \bottomrule
    \end{tabular}
    \caption{Agreement between a single blind human coder and the annotation pipeline, per
    act: Cohen's $\kappa$ with 95\% bootstrap intervals, and the number of positive labels
    each assigned. Participant acts use the blind recode of the first 65 turns.}
    \label{tab:human-kappa}
\end{table}

\end{document}